\documentclass[a4paper,11pt]{article}
\pdfoutput=1 

\usepackage{jinstpub}
\usepackage{subcaption}
\usepackage{tabularx}
\newcolumntype{C}{>{\centering\arraybackslash}X}
\usepackage{longtable}
\usepackage{arydshln}
\usepackage{float}
\usepackage{lineno}

\title{Scintillator ageing of the T2K near detectors from 2010 to 2021}

\affiliation[1]{University Autonoma Madrid, Department of Theoretical Physics, 28049 Madrid, Spain}
\affiliation[2]{University of Bern, Albert Einstein Center for Fundamental Physics, Laboratory for High Energy Physics (LHEP), Bern, Switzerland}
\affiliation[3]{Boston University, Department of Physics, Boston, Massachusetts, U.S.A.}
\affiliation[4]{University of California, Irvine, Department of Physics and Astronomy, Irvine, California, U.S.A.}
\affiliation[5]{IRFU, CEA, Universit\'e Paris-Saclay, F-91191 Gif-sur-Yvette, France}
\affiliation[6]{University of Colorado at Boulder, Department of Physics, Boulder, Colorado, U.S.A.}
\affiliation[7]{Colorado State University, Department of Physics, Fort Collins, Colorado, U.S.A.}
\affiliation[8]{Duke University, Department of Physics, Durham, North Carolina, U.S.A.}
\affiliation[9]{E\"{o}tv\"{o}s Lor\'{a}nd University, Department of Atomic Physics, Budapest, Hungary}
\affiliation[10]{ETH Zurich, Institute for Particle Physics and Astrophysics, Zurich, Switzerland}
\affiliation[11]{CERN European Organization for Nuclear Research, CH-1211 Gen\'eve 23, Switzerland}
\affiliation[12]{University of Geneva, Section de Physique, DPNC, Geneva, Switzerland}
\affiliation[13]{University of Glasgow, School of Physics and Astronomy, Glasgow, United Kingdom}
\affiliation[14]{H. Niewodniczanski Institute of Nuclear Physics PAN, Cracow, Poland}
\affiliation[15]{High Energy Accelerator Research Organization (KEK), Tsukuba, Ibaraki, Japan}
\affiliation[16]{University of Houston, Department of Physics, Houston, Texas, U.S.A.}
\affiliation[17]{Institut de Fisica d'Altes Energies (IFAE) - The Barcelona Institute of Science and Technology, Campus UAB, Bellaterra (Barcelona) Spain}
\affiliation[18]{Institut f\"ur Physik, Johannes Gutenberg-Universit\"at Mainz, Staudingerweg 7, 55128 Mainz, Germany}
\affiliation[19]{IFIC (CSIC \& University of Valencia), Valencia, Spain}
\affiliation[20]{Institute For Interdisciplinary Research in Science and Education (IFIRSE), ICISE, Quy Nhon, Vietnam}
\affiliation[21]{Imperial College London, Department of Physics, London, United Kingdom}
\affiliation[22]{INFN Sezione di Bari and Universit\`a e Politecnico di Bari, Dipartimento Interuniversitario di Fisica, Bari, Italy}
\affiliation[23]{INFN Sezione di Napoli and Universit\`a di Napoli, Dipartimento di Fisica, Napoli, Italy}
\affiliation[24]{INFN Sezione di Padova and Universit\`a di Padova, Dipartimento di Fisica, Padova, Italy}
\affiliation[25]{INFN Sezione di Roma and Universit\`a di Roma ``La Sapienza'', Roma, Italy}
\affiliation[26]{Institute for Nuclear Research of the Russian Academy of Sciences, Moscow, Russia}
\affiliation[27]{International Centre of Physics, Institute of Physics (IOP), Vietnam Academy of Science and Technology (VAST), 10 Dao Tan, Ba Dinh, Hanoi, Vietnam}
\affiliation[28]{International Laboratory for Astrophysics, Neutrino and Cosmology Experiments, Kashiwanoha, Japan}
\affiliation[29]{Kavli Institute for the Physics and Mathematics of the Universe (WPI), The University of Tokyo Institutes for Advanced Study, University of Tokyo, Kashiwa, Chiba, Japan}
\affiliation[30]{Keio University, Department of Physics, Kanagawa, Japan}
\affiliation[31]{King's College London, Department of Physics, Strand, London WC2R 2LS, United Kingdom}
\affiliation[32]{Kobe University, Kobe, Japan}
\affiliation[33]{Kyoto University, Department of Physics, Kyoto, Japan}
\affiliation[34]{Lancaster University, Physics Department, Lancaster, United Kingdom}
\affiliation[35]{Lawrence Berkeley National Laboratory, Berkeley, CA 94720, USA}
\affiliation[36]{Ecole Polytechnique, IN2P3-CNRS, Laboratoire Leprince-Ringuet, Palaiseau, France}
\affiliation[37]{University of Liverpool, Department of Physics, Liverpool, United Kingdom}
\affiliation[38]{Louisiana State University, Department of Physics and Astronomy, Baton Rouge, Louisiana, U.S.A.}
\affiliation[39]{Joint Institute for Nuclear Research, Dubna, Moscow Region, Russia}
\affiliation[40]{Michigan State University, Department of Physics and Astronomy,  East Lansing, Michigan, U.S.A.}
\affiliation[41]{Miyagi University of Education, Department of Physics, Sendai, Japan}
\affiliation[42]{National Centre for Nuclear Research, Warsaw, Poland}
\affiliation[43]{State University of New York at Stony Brook, Department of Physics and Astronomy, Stony Brook, New York, U.S.A.}
\affiliation[44]{Okayama University, Department of Physics, Okayama, Japan}
\affiliation[45]{Osaka City University, Department of Physics, Osaka, Japan}
\affiliation[46]{Oxford University, Department of Physics, Oxford, United Kingdom}
\affiliation[47]{University of Pennsylvania, Department of Physics and Astronomy,  Philadelphia, PA, 19104, USA.}
\affiliation[48]{University of Pittsburgh, Department of Physics and Astronomy, Pittsburgh, Pennsylvania, U.S.A.}
\affiliation[49]{Queen Mary University of London, School of Physics and Astronomy, London, United Kingdom}
\affiliation[50]{University of Regina, Department of Physics, Regina, Saskatchewan, Canada}
\affiliation[51]{University of Rochester, Department of Physics and Astronomy, Rochester, New York, U.S.A.}
\affiliation[52]{Royal Holloway University of London, Department of Physics, Egham, Surrey, United Kingdom}
\affiliation[53]{RWTH Aachen University, III. Physikalisches Institut, Aachen, Germany}
\affiliation[54]{Departamento de F\'isica At\'omica, Molecular y Nuclear, Universidad de Sevilla, 41080 Sevilla, Spain}
\affiliation[55]{University of Sheffield, Department of Physics and Astronomy, Sheffield, United Kingdom}
\affiliation[56]{University of Silesia, Institute of Physics, Katowice, Poland}
\affiliation[57]{Sorbonne Universit\'e, Universit\'e Paris Diderot, CNRS/IN2P3, Laboratoire de Physique Nucl\'eaire et de Hautes Energies (LPNHE), Paris, France}
\affiliation[58]{STFC, Rutherford Appleton Laboratory, Harwell Oxford,  and  Daresbury Laboratory, Warrington, United Kingdom}
\affiliation[59]{University of Tokyo, Department of Physics, Tokyo, Japan}
\affiliation[60]{University of Tokyo, Institute for Cosmic Ray Research, Kamioka Observatory, Kamioka, Japan}
\affiliation[61]{University of Tokyo, Institute for Cosmic Ray Research, Research Center for Cosmic Neutrinos, Kashiwa, Japan}
\affiliation[62]{Tokyo Institute of Technology, Department of Physics, Tokyo, Japan}
\affiliation[63]{Tokyo Metropolitan University, Department of Physics, Tokyo, Japan}
\affiliation[64]{Tokyo University of Science, Faculty of Science and Technology, Department of Physics, Noda, Chiba, Japan}
\affiliation[65]{University of Toronto, Department of Physics, Toronto, Ontario, Canada}
\affiliation[66]{TRIUMF, Vancouver, British Columbia, Canada}
\affiliation[67]{University of Warsaw, Faculty of Physics, Warsaw, Poland}
\affiliation[68]{Warsaw University of Technology, Institute of Radioelectronics and Multimedia Technology, Warsaw, Poland}
\affiliation[69]{Tohoku University, Faculty of Science, Department of Physics, Miyagi, Japan}
\affiliation[70]{University of Warwick, Department of Physics, Coventry, United Kingdom}
\affiliation[71]{University of Winnipeg, Department of Physics, Winnipeg, Manitoba, Canada}
\affiliation[72]{Wroclaw University, Faculty of Physics and Astronomy, Wroclaw, Poland}
\affiliation[73]{Yokohama National University, Department of Physics, Yokohama, Japan}
\affiliation[74]{York University, Department of Physics and Astronomy, Toronto, Ontario, Canada}
\author[60]{K.\,Abe,}
\author[49]{N.\,Akhlaq,}
\author[66]{R.\,Akutsu,}
\author[71,66]{A.\,Ali,}
\author[10]{C.\,Alt,}
\author[58,37]{C.\,Andreopoulos,}
\author[19]{M.\,Antonova,}
\author[32]{S.\,Aoki,}
\author[63]{T.\,Arihara,}
\author[73]{Y.\,Asada,}
% ...

\abstract{The T2K experiment widely uses plastic scintillator as a target for neutrino interactions and an active medium for the measurement of charged particles produced in neutrino interactions at its near detector complex. Over 10 years of operation the measured light yield recorded by the scintillator based subsystems has been observed to degrade by 0.9--2.2\% per year. Extrapolation of the degradation rate through to 2040 indicates the recorded light yield should remain above the lower threshold used by the current reconstruction algorithms for all subsystems. This will allow the near detectors to continue contributing to important physics measurements during the T2K-II and Hyper-Kamiokande eras. Additionally, work to disentangle the degradation of the plastic scintillator and wavelength shifting fibres shows that the reduction in light yield can be attributed to the ageing of the plastic scintillator.}


\collaboration{\includegraphics[height=17mm]{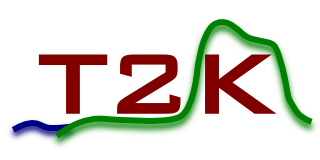}\\[6pt]
  The T2K Collaboration}

\begin{document}
\maketitle
\flushbottom

\section{Introduction} \label{sec:intro}

\subsection{The T2K Experiment} \label{sec:T2K}

T2K (Tokai-to-Kamioka) is a long-baseline neutrino oscillation experiment \cite{ABE2011106} located in Japan, measuring muon (anti-)neutrino disappearance and electron (anti-)neutrino appearance from a muon (anti-)neutrino beam produced by the J-PARC (Japan Proton Accelerator Research Complex) synchrotron \cite{10.1093/ptep/pts071}.
The experiment consists of  a far detector at a distance of 295\,km from J-PARC, a near detector complex 280\,m downstream of the proton beam target, and the beam facility itself.
The far detector is Super-Kamiokande \cite{FUKUDA2003418}, a 50\,kt water Cherenkov detector positioned $2.5^{\circ}$ off the beam axis.

The near detector complex contains  ND280 \cite{ND280TDR} and INGRID \cite{Abe:2011xv} detectors which started operation in 2010. In addition, WAGASCI-BabyMIND \cite{wagasci, Baby_mind} was installed in 2019.
INGRID is located directly on the beam axis, while ND280 is situated at the same off-axis angle as Super-Kamiokande.
ND280 measures the rate of neutrino interactions before oscillation has occurred.  This provides information on the neutrino flux, cross section and neutrino type which is necessary to predict the interaction rate at the far detector.
INGRID monitors the neutrino beam direction and profile as well as the neutrino interaction event rate with high statistics.

Most subsystems of  ND280, along with the INGRID detector, use plastic scintillator bars as an active detector medium.
Whilst traversing the detector, charged particles excite electrons within the scintillator material to higher orbitals. The de-excitation of the electrons produces the emission of scintillation light which is used to track the passage of these particles.
The scintillation light is collected by 1\,mm diameter Kuraray wavelength-shifting (WLS) fibres \cite{WLS} for transmission to Hamamatsu Multi-Pixel Photon Counters (MPPC) \cite{YOKOYAMA2009128}, a type of Silicon Photon Multiplier (SiPM) located at one or both ends of the scintillator bars.

T2K was the first experiment to employ MPPCs on a large scale, utilising $\sim$65,000 MPPCs across the near detectors. The observed MPPC failure rate has been very low, at around $\sim$0.5\% of the total over the current lifetime of the experiment, and so their failure is not currently a concern for the future operation of the T2K near detectors.

However, during 10 years of operation some degradation in the light yield produced by the scintillator bars has occurred. Similar degradation has also been observed in other experiments, such as MINOS \cite{Michael:2008bc} and MINER$\nu$A \cite{Aliaga:2013uqz}. Understanding this effect is important for the accurate calibration of the detectors, for monitoring their long-term efficiency and predicting the future performance.

\subsection{Scintillator Ageing} \label{sec:scintageing}

The issue of plastic scintillator ageing is long known \cite{osti_4793399}, and there are many studies aimed at measuring, characterizing and developing stabilisation methods for these widely used materials (see for example \cite{osti_10158865, ARTIKOV2005125, zhouhesheng, osti_1662050, osti_1356922, loyd, ZAITSEVA2020161709}). These studies often consider the impact of potentially controllable environmental factors such as temperature and humidity on the long-term performance of the materials, as well as ways to chemically stabilise them.

The exact mechanism for scintillator ageing occurring within the T2K near detectors is unknown, but there are a number of potentially contributing factors:
\begin{itemize}
    \item Mechanical stressing of the scintillator causing the development of crazes or shears within the material \cite{kambour}. These inhibit the uniform scattering of light within the scintillator, preventing transmission through total internal reflection.
    \item Fogging of the scintillators due to water penetrating into the material and condensing \cite{7027257}. This increases the opacity of the scintillator and is a significant problem where the materials are exposed to very high humidity conditions with large temperature variations.
    \item Oxidation of the scintillator through photochemical processes that lead to the production of peroxides causing the yellowing of the material \cite{pmid25674392}. This reduces the light yield from the scintillator and has been observed in the accelerated ageing test performed on the scintillator bars used by the MINOS experiment \cite{Michael:2008bc}, which are materially identical to the INGRID, FGD, ECal and P\O{}D subsystems of T2K as described in section \ref{sec:ND280}.
\end{itemize}

Within this paper the relevant T2K near detector subsystems are described in section \ref{sec:T2KDets} and the data samples and light yield measurement methods used are detailed in section \ref{sec:ly_measuremets}. The rate of degradation of the T2K scintillator is presented in section \ref{sec:ageing}, along with predictions for the future response of the detectors in section \ref{sec:futureresponse} and an attempt to disentangle whether the ageing is dominated by the degradation of the scintillator or wavelength shifting (WLS) fibres in section \ref{sec:scintandfibreageing}.

\section{The T2K Scintillator Detectors} \label{sec:T2KDets}

\subsection{ND280} \label{sec:ND280}
\begin{figure}[t]
    \centering
    \includegraphics[scale=0.5]{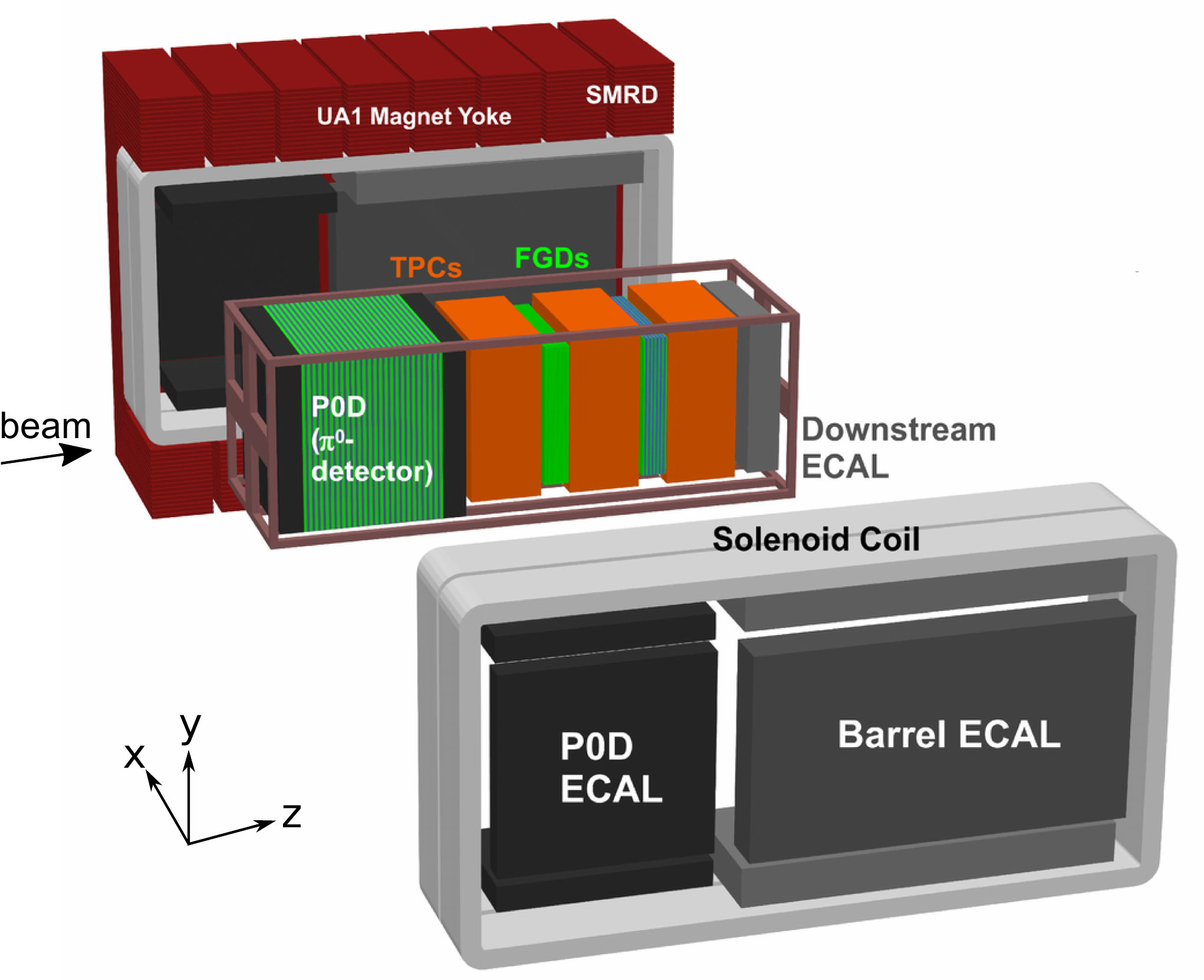}
    \caption{Exploded diagram of the ND280 off-axis near detector displaying the different detector subsystems.}
    \label{fig:nd280}
\end{figure}

The ND280 detector, figure~\ref{fig:nd280}, is composed of a set of subsystems enclosed within the refurbished UA1 magnet~\cite{Timmer:1983za}. The subsystems are as follows:
\begin{itemize}
    \item A detector composed of scintillator, water and brass target planes designed to identify $\pi^{0}$s (P\O D) \cite{ASSYLBEKOV201248}.
    \item The tracker region, consisting of three time projection chambers (TPCs) \cite{ABGRALL201125} and two plastic scintillator fine-grained detectors (FGDs) \cite{Amaudruz:2012esa}, optimised to study charged current interactions of incoming neutrinos. The upstream FGD1 is entirely composed of scintillator planes, the downstream FGD2 consists of alternating modules of scintillator and water-filled volumes.  
    \item Plastic scintillator and lead sampling electromagnetic calorimeters (ECals) that surround the P\O D and tracker region \cite{Allan:2013ofa}.
    \item Plastic scintillator side muon range detectors (SMRDs) \cite{Aoki:2012mf} situated in the magnet flux return yokes.
\end{itemize}
The coordinate system has $z$ along the neutrino beam direction, and $x$ and $y$ are horizontal and vertical, respectively.

The material used in the P\O D, FGD and ECal scintillator bars is polystyrene Dow Styron 663\,W, doped with 1\% PPO (2,5-diphenyloxazole) and 0.03\% POPOP (1,4-bis(5-phenyloxazol-2-yl) benzene) and co-extruded with a surface layer of polystyrene loaded with 15\% TiO$_2$ to allow diffuse reflection of scintillation light.
The bars for the P\O D and ECal were manufactured in the extrusion facility at Fermi National Accelerator Laboratory (FNAL) and their composition is identical to that of the scintillator bars used in the MINOS experiment \cite{Michael:2008bc}.
For the FGD, scintillator bars of the same composition were produced by extrusion procedure at Celco Plastics Ltd, Surrey, British Columbia.
The scintillator bars of the SMRD use polystyrene doped with 1.5\% PTP (1,4-Diphenylbenzene) and 0.01\% POPOP and were chemically etched to produce a reflective coating. These were manufactured by the Uniplast company in Vladimir, Russia. All the ND280 scintillator bars were produced between 2006 and 2009 as shown in table \ref{tab:production_dates}.
\begin{table}[t]
    \footnotesize
    \centering
    \caption{Scintillator production dates for each detector.}
    \begin{tabular}{|c|c|}
    \hline
    Detector & Production Period \\
    \hline
    P\O D & 2007--2008 \\
    ECal  & 2007--2009 \\
    FGD   & 2006 \\
    SMRD  & 2007--2008 \\
    \hdashline
    INGRID & 2007--2008 \\
    \hline 
    \end{tabular}
    \label{tab:production_dates}
\end{table}

All subsystems use Kuraray Y\nobreakdash-11 blue to green WLS fibres for photon transmission to the MPPCs. The specific WLS formulations used are Y\nobreakdash-11(175) S-35 J-type (P\O D), Y\nobreakdash-11(200) S-35 J-type (FGD), Y\nobreakdash-11(200) CS-35 J-type (ECal) and Y\nobreakdash-11(150) S-70 S-type (SMRD).

The particular geometry of each subsystem is described below.

\subsubsection{P\O D}
\begin{figure}[t]
    \centering
    \includegraphics[scale=1.0]{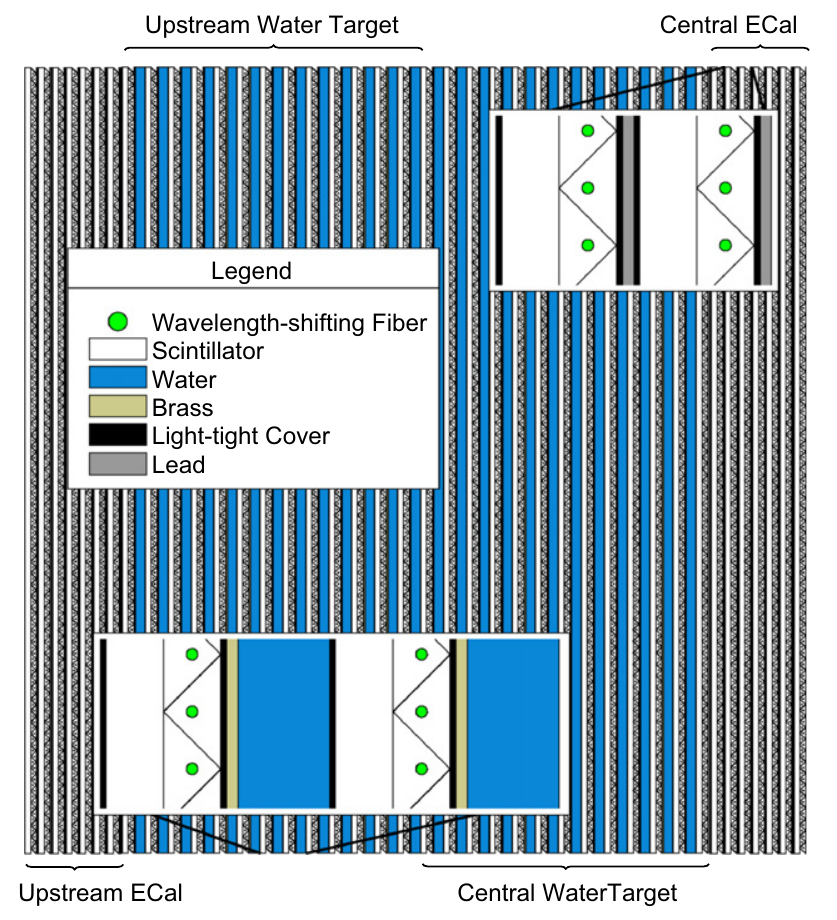}
    \caption{A schematic of the four PØD Super-PØDules as installed in the detector. The neutrino beam enters from the left hand side of the figure.}
    \label{fig:p0d_schematic}
\end{figure}

\begin{figure}[t]
\center
\begin{subfigure}{0.49\textwidth}
    \center{\includegraphics[width=\linewidth]{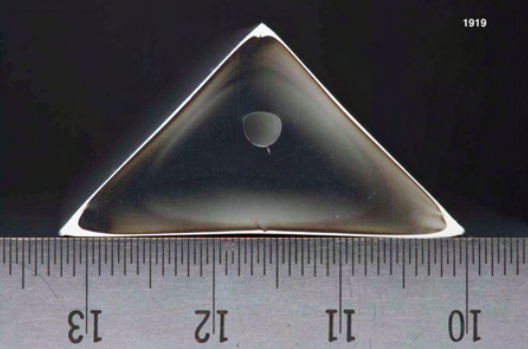}}
    \caption{End view of a P\O D scintillator bar.}
    \label{fig:p0d_bar_image}
\end{subfigure}
\hfill
\begin{subfigure}{0.49\textwidth}
    \center{\includegraphics[width=\linewidth]{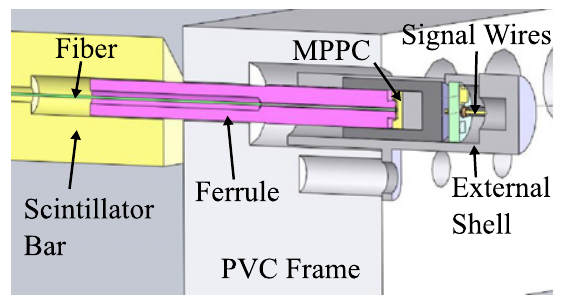}}
    \caption{A close-up view of the edge of a PØDule showing how the WLS fibers exit the scintillator bars and couple to the MPPCs.}
    \label{fig:p0d_mppc_connection_schematic}
\end{subfigure}
\caption{P0D bar image (a) and MPPC connection schematic (b).}
\label{fig:p0d_bar_pics}
\end{figure}

The P\O D detector consists of 40 scintillator modules called P\O Dules, as shown in~figure~\ref{fig:p0d_schematic}.
Each P\O Dule consists of two orthogonally oriented bar layers sandwiched between an inactive target and radiator material. The P\O Dules are  perpendicular to the beam direction and are assembled into four constituent units called Super-P\O Dules, these are defined in the following way:
\begin{enumerate}
 \item Super-P\O Dule 0: P\O Dules 0 - 6, the Upstream ECal.
 \item Super-P\O Dule 1: P\O Dules 7 - 19, the Upstream Water Target.
 \item Super-P\O Dule 2: P\O Dules 20 - 32, the Central Water Target. 
 \item Super-P\O Dule 3: P\O Dules 33 - 39, the Central ECal.
\end{enumerate}
The scintillator bars in the P\O D are triangular in cross section with a height of 17\,mm and a width of 33\,mm. Each bar has a single 2.6\,mm diameter coaxial hole through which a WLS fibre is inserted, as shown in figure~\ref{fig:p0d_bar_image}.
The horizontal bars are 2133\,mm long and the vertical bars are 2272\,mm long.
The fibres are not secured within the bar, leaving an air gap between the bar and WLS fibre.
The WLS fibres are mirrored with a vacuum deposition of aluminium on one end and are optically coupled to an MPPC on the other.
A ferrule is glued to the end of the fibre which couples to a housing holding the MPPC, see figure~\ref{fig:p0d_mppc_connection_schematic}.
A 3 mm thick polyethylene disk behind the MPPC provides pressure between the fibre and MPPC epoxy window. This design of fibre to MPPC coupling is also used by the ECal (see section  \ref{sec:nd280ecal}).

\subsubsection{FGD}
\begin{figure}[t]
\center
\begin{subfigure}{0.35\textwidth}
    \center{\includegraphics[width=\linewidth]{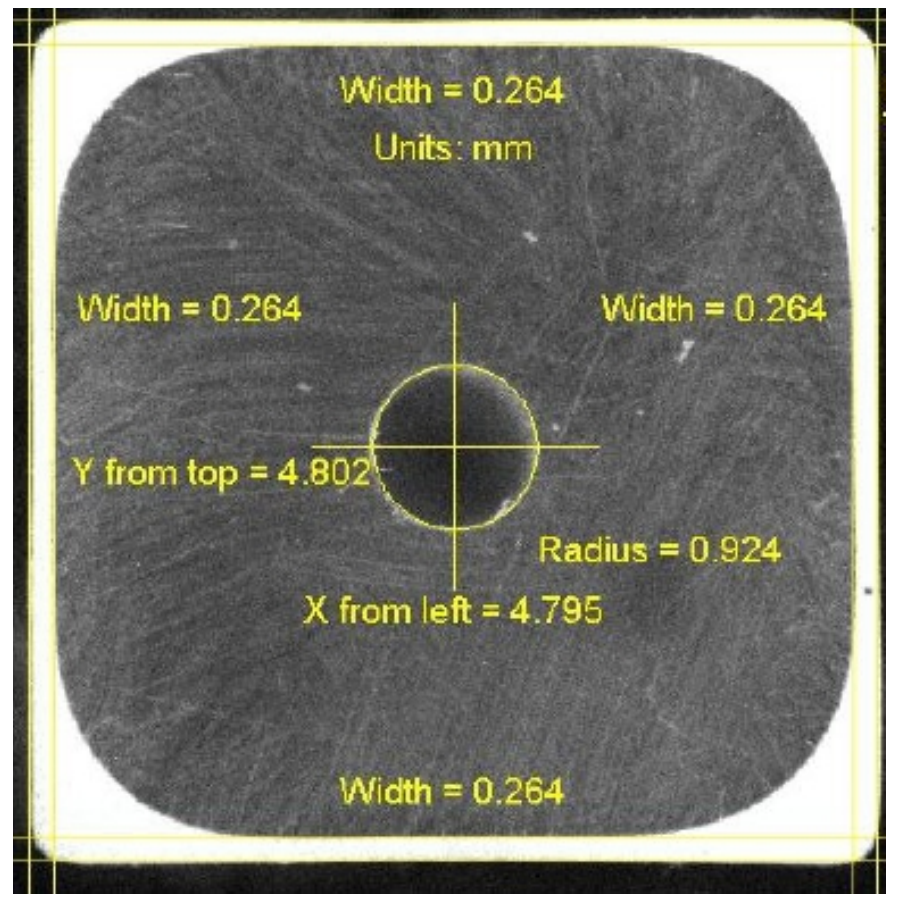}}
    \caption{Photo  taken  with  a  CCD  camera  of  a  typical FGD scintillator bar.}
    \label{fig:fgd_bar}
\end{subfigure}
\hfill
\begin{subfigure}{0.49\textwidth}
    \center{\includegraphics[width=\linewidth]{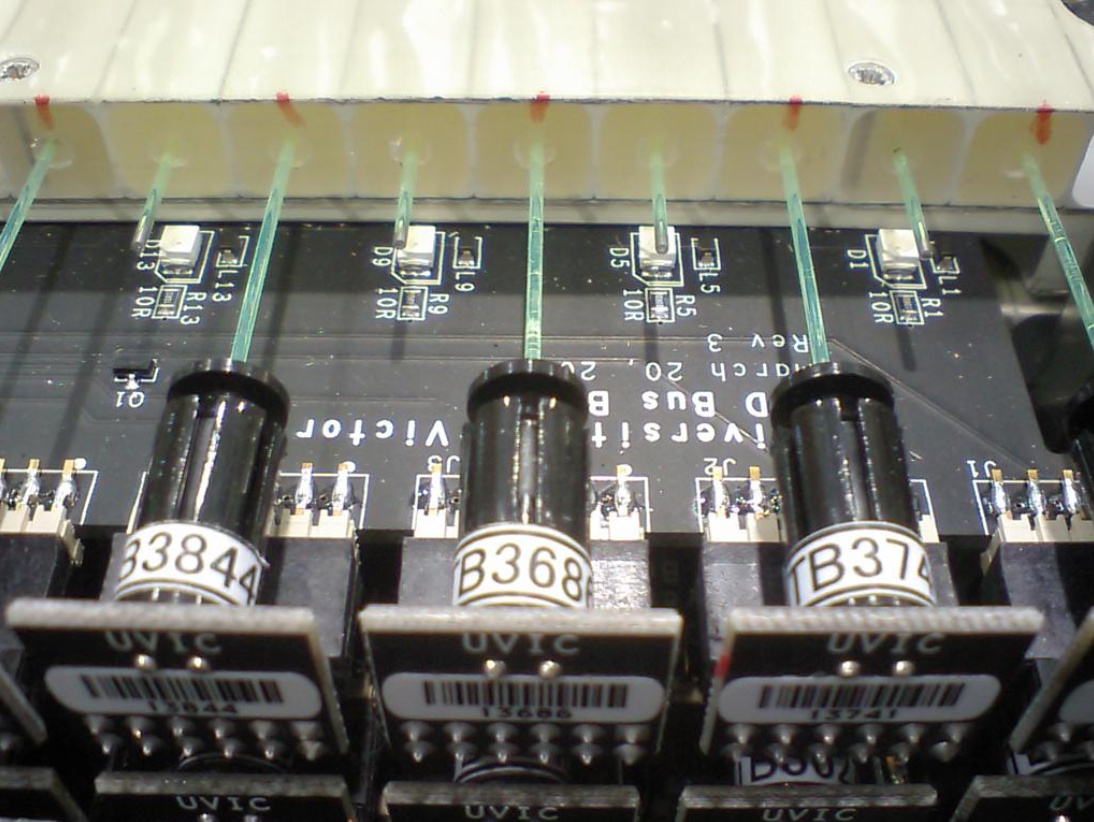}}
    \caption{Partial view of the end of a FGD scintillator layer with alternating fibres connected to MPPCs.}
    \label{fig:fgd_layer_image}
\end{subfigure}
\caption{An image of a FGD scintillator bar (a) and a FGD scintillator layer (b).}
\label{fig:fgd_bar_pics}
\end{figure}

The FGD scintillator bars are perpendicular to the beam in either the horizontal (X) or vertical (Y) direction, and are arranged into "XY" modules.
Each module consists of a layer of 192 bars in the horizontal direction glued to 192 bars in the vertical direction.
The scintillator bars have a square cross section with a side width of 9.6\,mm.
The length of the bars is 1864\,mm and each has a 1.8\,mm diameter hole through its centre containing the WLS fibre, see figure~\ref{fig:fgd_bar}.
One end of the fibre is mirrored with a vacuum deposition of aluminium to improve light collection efficiency, the other end is connected to an MPPC.
The upstream FGD1 contains fifteen such modules while the downstream FGD2 contains seven modules interspersed with inactive water target layers.
Each FGD module has dimensions of $1864 \times 1864 \times 20.2$\,mm\textsuperscript{3} (not including electronics).
There is an air gap between the scintillator and the fibre.
The fibre extends only a few centimetres from one end of the bar to reach an MPPC as shown in figure~\ref{fig:fgd_layer_image}. The fibre is connected to the MPPC with a custom two part connector, one part glued to the fibre and the other holding the MPPC, latched together by mechanical force. Bicron BC600 glue was chosen to fix the coupler to the fiber.  
Within each layer, alternate bars are read out from alternating ends.

\subsubsection{ECal} \label{sec:nd280ecal}
\begin{figure}[t]
    \center
    \begin{subfigure}{0.49\textwidth}
    \includegraphics[width=\linewidth]{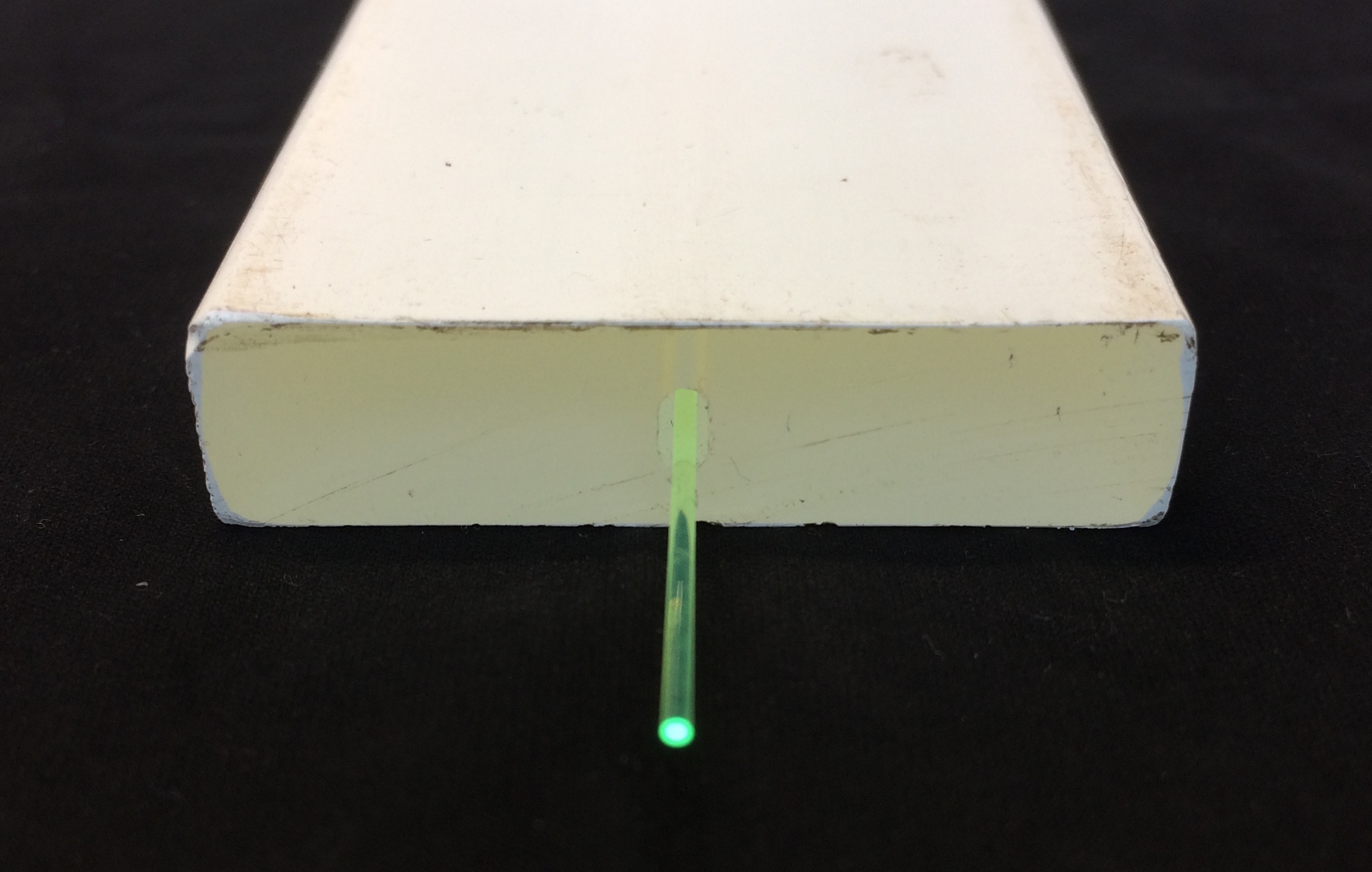}
    \caption{End view of an ECal scintillator bar and WLS fibre.}
    \label{fig:ecal_bar}
    \end{subfigure}
    \hfill
    \begin{subfigure}{0.45\textwidth}
    \includegraphics[width=\linewidth]{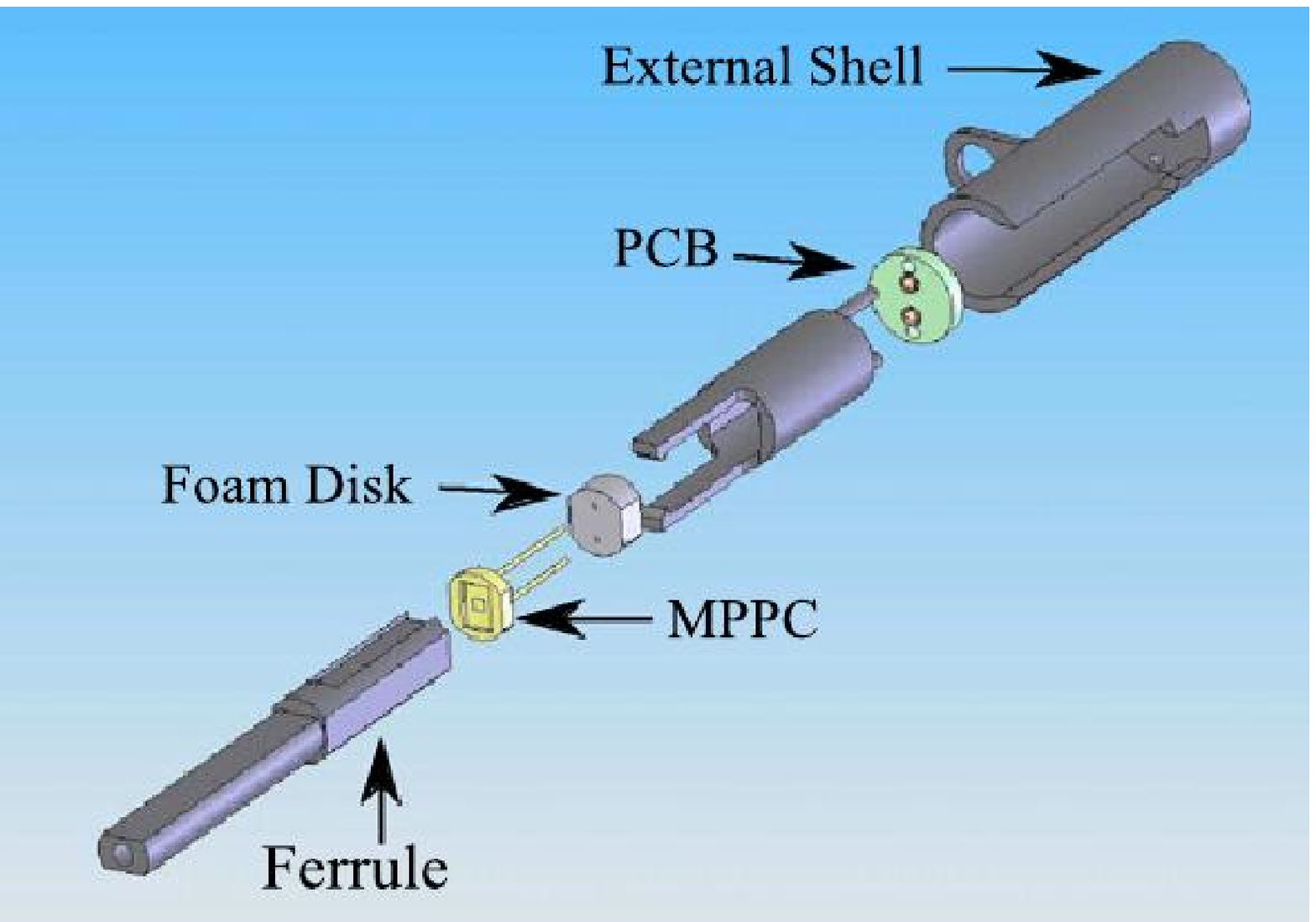}
    \caption{Exploded diagram of WLS fibre to MPPC connector.}
    \label{fig:ecal_connect}
    \end{subfigure}
    \caption{ECal scintillator bar with WLS fibre (a) and diagram of the WLS fibre to MPPC connector assembly (b).}
\end{figure}

The ECal scintillator bars have a cross section of $40\times10$~mm\textsuperscript{2} with a 2\,mm diameter hole down the centre through which a WLS fibre passes, see figure~\ref{fig:ecal_bar}. There is an air gap between the scintillator and the fibre.
The ECal is comprised of thirteen modules each of which uses one or two different lengths of scintillator bar in their construction.
The Downstream (DS) ECal module has 1700 bars of length 2000\,mm oriented in alternating vertical and horizontal layers perpendicular to the beam direction.
Across the six Barrel ECal modules there are 3990 bars of length 3840\,mm (Z bars) lying parallel to the beam direction. The four top and bottom barrel modules contain 6144 bars of length 1520\,mm (X bars), and the two side bbarrel modules contain 3072 bars of length 2280\,mm (Y bars).
The Barrel X and Y bars are oriented perpendicular to the beam direction and have fibres which are mirrored on one end with a vacuum deposition of aluminium, while the other end is connected to an MPPC as shown in figure \ref{fig:ecal_connect}.
The Barrel Z and Downstream ECal bars have fibres which are connected to MPPCs on both ends.
As will be described in section \ref{sec:ly_measuremets}, the analysis methods used within this paper require the 3D reconstruction of muon tracks. As a result the six P\O D ECal modules (not described) are not used in the studies presented as all the scintillator bars run parallel to the beam direction making the 3D reconstruction of particle tracks, needed for ageing studies, challenging.
It should be noted that the Downstream ECal was installed into ND280 in early-2010, while the Barrel ECal modules were installed in late-2010.

\subsubsection{SMRD}
\begin{figure}[t]
\center
\begin{subfigure}{0.44\textwidth}
    \center{\includegraphics[width=\textwidth]{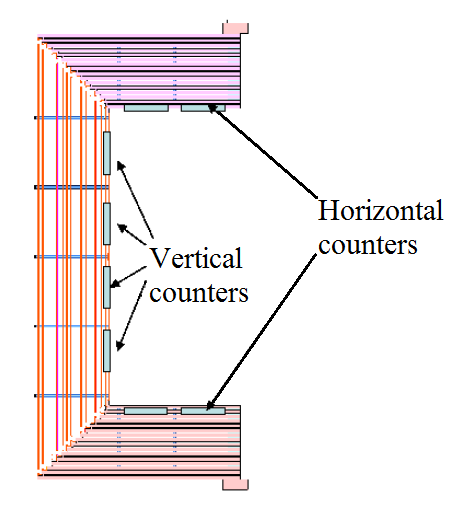}}
    \caption{Schematic view of a layer of SMRD modules.}
    \label{fig:smrd_schematic}
\end{subfigure}
\hfill
\begin{subfigure}{0.49\textwidth}
    \center{\includegraphics[width=\textwidth]{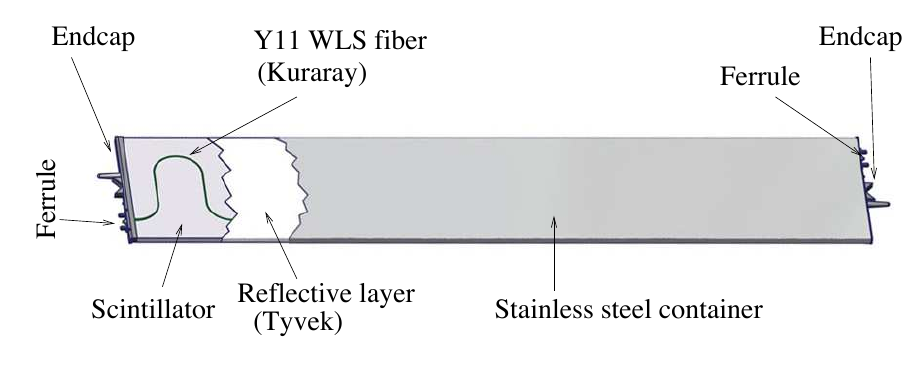}}
    \caption{SMRD counter sliced view.}
    \label{fig:smrd_counter_schematic}
\end{subfigure}
\caption{Schematic view of SMRD module positions (a) and image of an SMRD counter design (b).}
\label{fig:smrd_bars}
\end{figure}

There are two types of SMRD scintillator bars with different sizes, horizontal ($7 \times 167 \times 875$\,mm\textsuperscript{3}) and  vertical ($7 \times 175 \times 875$\,mm\textsuperscript{3}), grouped in modules of 4 and 5 respectively, see figure~\ref{fig:smrd_schematic}.
There are 404 modules in total with 768 (1240) horizontal (vertical) bars. 
The modules are placed in layers in the air gaps of the magnetic flux return yokes.
The magnet yokes are numbered 1-8 going downstream along the beam direction. 
All yokes host three horizontal layers and yokes 1 through 5 also host three vertical layers. The most downstream yokes host more vertical layers: yoke 6 hosts four, and yokes 7 and 8 host six.
Figure \ref{fig:smrd_schematic} shows the placement of the first layer of modules in a yoke segment.
For better collection of the scintillator light and to improve the positional accuracy in the SMRD,
S-shaped (curvature of $\varnothing=58$\,mm) WLS fibres run down the length or the bars as shown in figure~\ref{fig:smrd_counter_schematic}.
The fibres are bent and glued into grooves within the scintillator bars using BC600 Bicron glue. It is worth noting that any degradation of this glue with time could have an impact on the light yield measured by the SMRD.
The design results in nearly uniform response across the surface while reducing the number of channels to read out.
The signal is read out from both ends of the bar via MPPCs.
Each fibre exits through a ferrule which is part of a custom made endcap, glued and screwed to the scintillator, to which a connector with the MPPC is attached.
A foam spring ensures a reliable coupling between the photosensor and the fibre.

\subsection{INGRID}
\begin{figure}[t]
\centering
\begin{subfigure}{0.48\textwidth}
    \includegraphics[scale=0.54,clip]{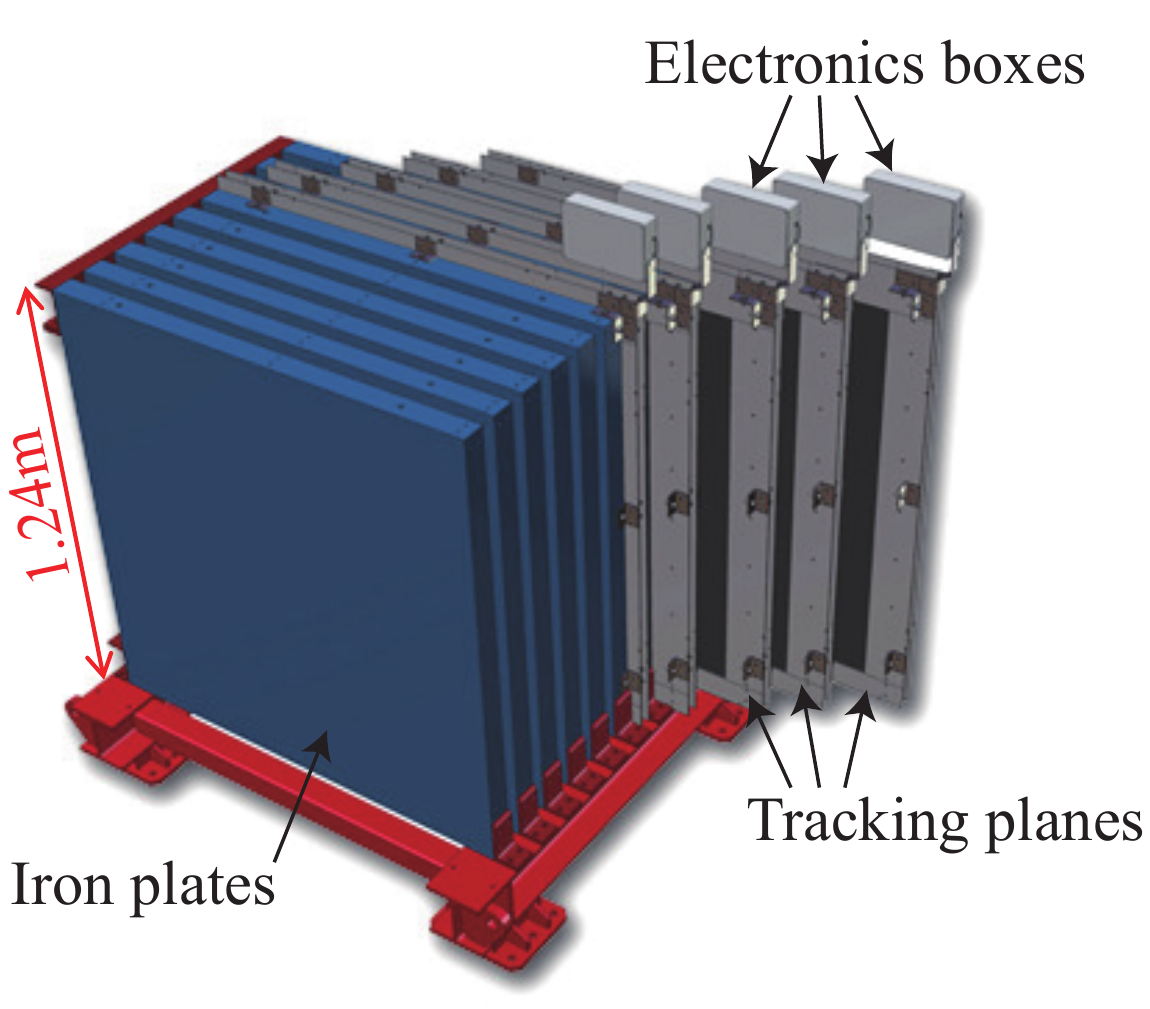}
    \caption{Iron plates with scintillator planes being inserted.}
\end{subfigure}
~
\begin{subfigure}{0.48\textwidth}
    \includegraphics[scale=0.5,clip]{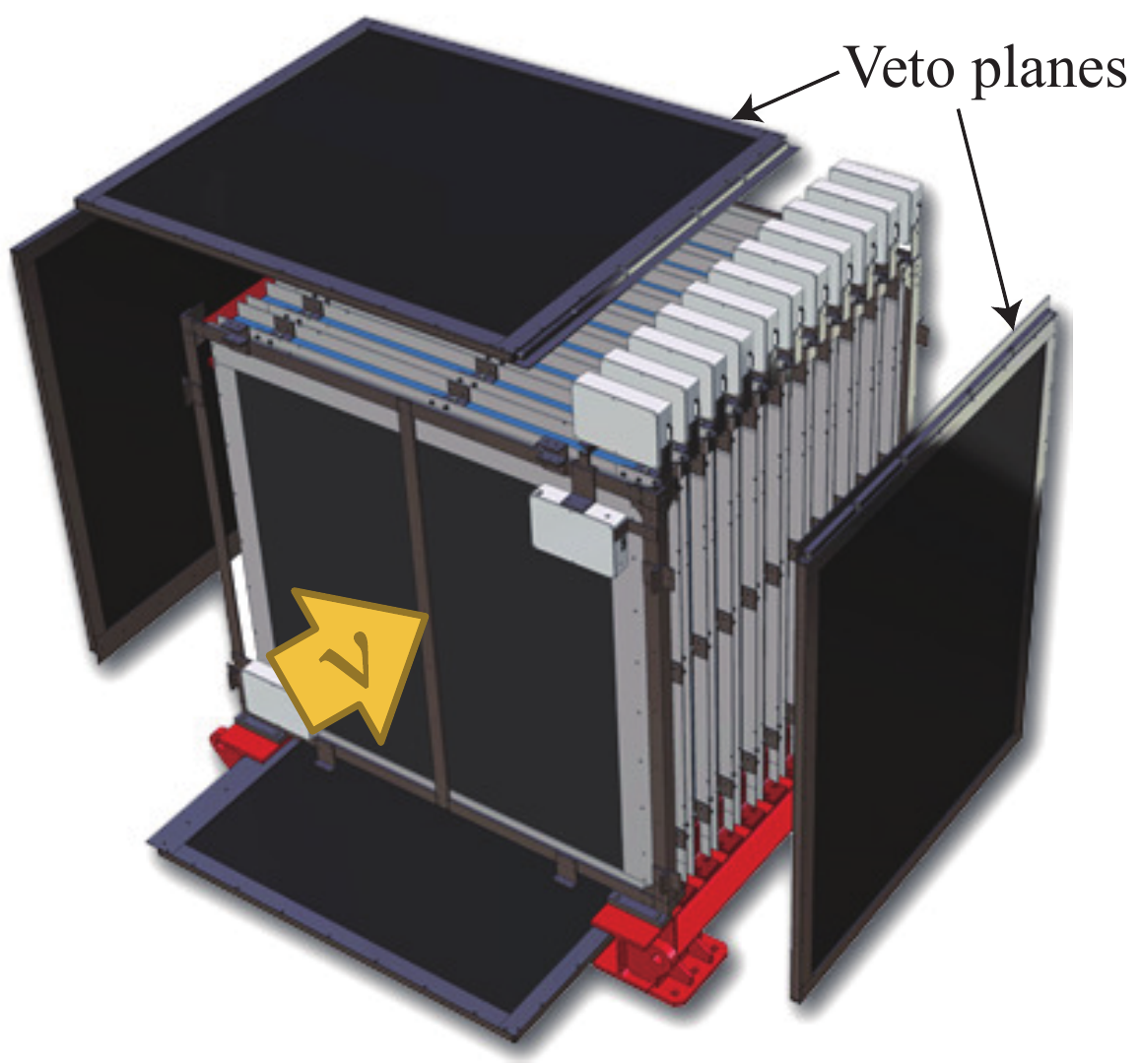}
    \caption{Full module structure with surrounding veto planes.}
\end{subfigure}
\caption{Structure of an INGRID module. Scintillator tracking planes are interleaved with iron plates (a). The sides of the module are then surrounded by scintillator veto planes (b).}
\label{fig:ingrid}
\end{figure}

The INGRID detector consists of 16 identical iron and plastic scintillator detector modules. Each module is constructed of 11 tracking plastic scintillator planes interleaved with 9 passive iron plates, as shown in figure~\ref{fig:ingrid} (the final pair of scintillator planes lacks an interleaved iron plate).
Each scintillator plane lies perpendicular to the beam direction and consists of 24 horizontally (X) orientated bars glued to a further 24 vertically (Y) orientated bars. Each bar is 1203\,mm long and has a cross section of $50\times10$\,mm$^2$.

\begin{figure}[t]
    \centering
    \includegraphics[scale=0.5]{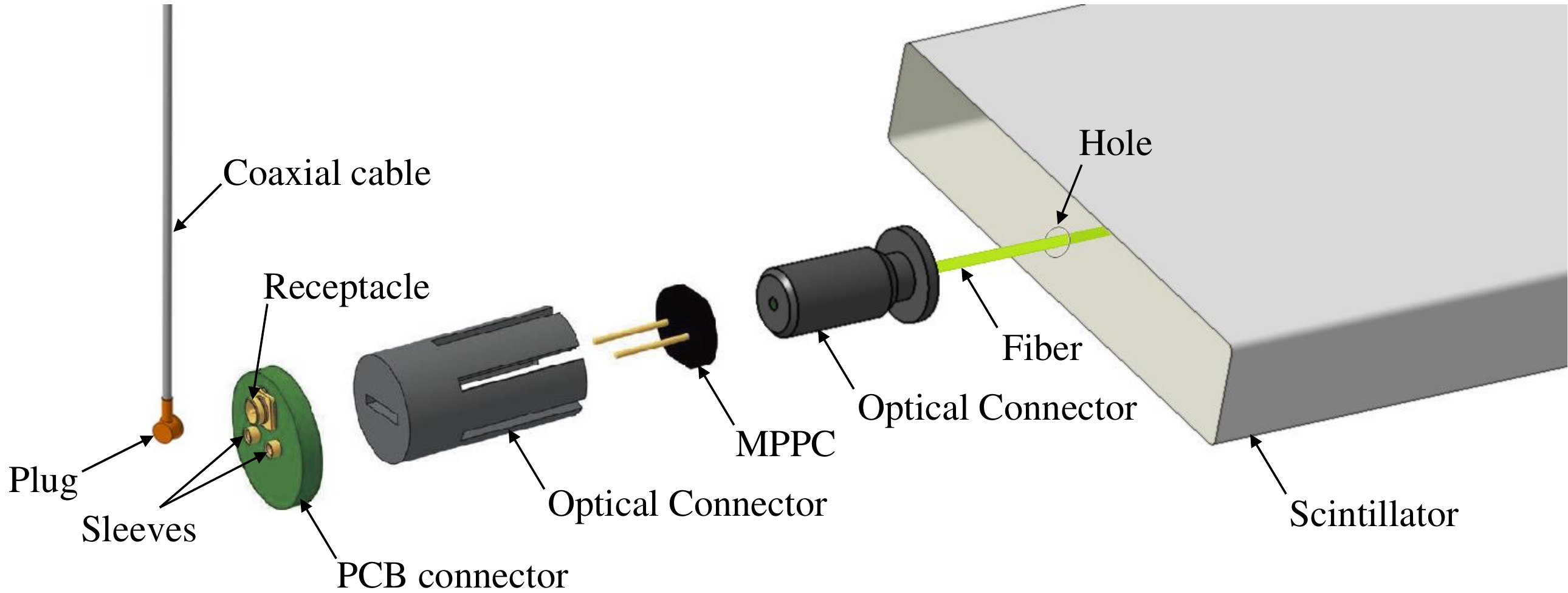}
    \caption{Schematic view of the readout components for INGRID.}
    \label{fig:ingrid_readout}
\end{figure}

In common with the P\O D and ECal scintillator bars, the INGRID bars were produced at Fermilab in 2007--2008. As such they have the same material composition of Dow Styron 663 W polystyrene, doped with 1\% PPO and 0.03\% POPOP, and co-extruded with a TiO$_2$ rich material to allow diffuse reflection of scintillation light. Unglued Kuraray Y11(200) M-type WLS fibres collect the light from the bars and are coupled on one end to MPPCs as shown in figure~\ref{fig:ingrid_readout}. The uninstrumented ends of the bars and fibres are painted with a reflective coating of ELJEN\textsuperscript{\textregistered} EJ-510. The design of the fibre-MPPC coupling for INGRID is the same as used in the FGDs.
\section{Light Yield Measurements} \label{sec:ly_measuremets}
The degradation in scintillator response can be quantified by measuring the change over time of the average light yield observed from the passage of minimum ionizing particles (MIP), through the ND280 and INGRID subsystems. The recorded and calibrated response of an MPPC due to the passage of a MIP through a scintillator bar constitutes a ``hit'' within a subsystem.

Due to the varying geometry and acceptance of the subsystems, several different MIP samples were used for the analysis: beam neutrino interactions, cosmic ray muons recorded concurrently with each T2K Run, or sand muon data (muons produced in neutrino interactions upstream of the detectors).

In all cases the MIP light yield was corrected to account for the length of the MIP's path through the scintillator bar based on the track angle, and attenuation in the WLS fibre based on the reconstructed position along the scintillator bar.

Regular ($\sim$weekly) adjustments were made to the MPPC overvoltage to account for temperature variations in order to maintain a stable gain, and therefore detector response, over time. This was achieved by stepping through a range of bias voltages and measuring the difference (gain) between the pedestal and single photoelectron response for each MPPC. The correlation between measured gain as a function of voltage was used to extract the appropriate overvoltage to be applied to each MPPC. For ND280 this is supplemented by more frequent calibrations ($\sim$3 hourly) that are derived from the recorded detector temperature (FGD) or directly from the pedestal and single photoelectron response of the MPPCs (ECal, SMRD, P\O{}D) and applied during offline reconstruction. Additional empirically derived corrections were also applied to account for the changes in photodetection efficiency, cross talk and after-pulsing as a function of overvoltage \cite{Amaudruz:2012esa,Allan:2013ofa}. INGRID only uses the pedestal and gain measured after the weekly MPPC bias voltage adjustments for their calibration without additional offline fine-tuning.

\subsection{Data Samples}
\begin{figure}[t]
    \centering
    \includegraphics[width=0.8\linewidth]{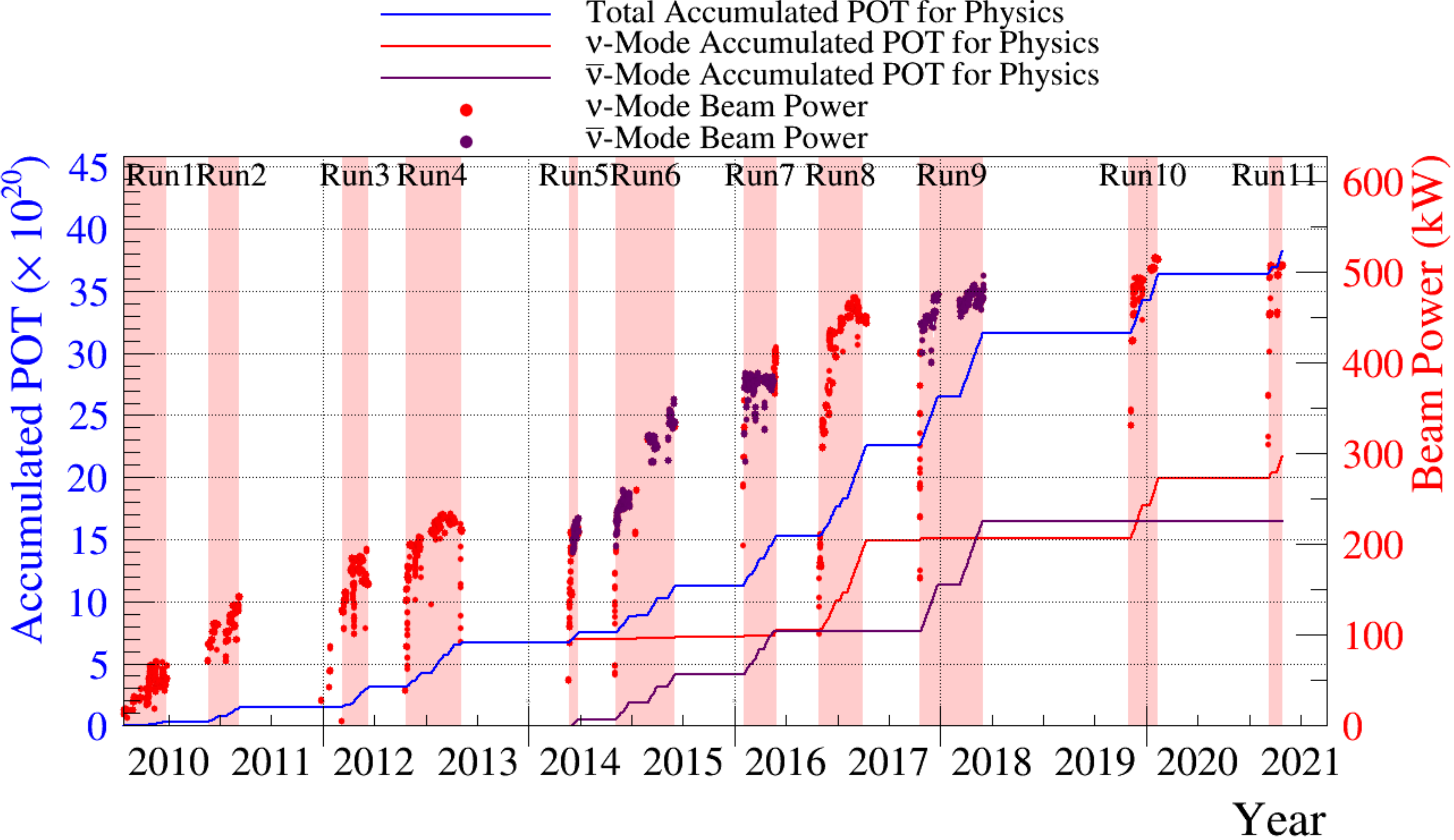}
    \caption{T2K Run periods, associated beam power and accumulated protons on target (POT).}
    \label{fig:accumPOT}
\end{figure}

T2K first became operational in March 2010 and neutrino beam data had been recorded during 11 separate T2K Run periods by the end of 2021, as shown in figure~\ref{fig:accumPOT} and table~\ref{tab:t2kruns}. Data taken during T2K Runs 1--11 and 1--9 were used by the INGRID and ND280 subsystems, respectively, in the analyses described by this paper.
\begin{table}[t]
    \footnotesize
    \centering
    \caption{Dates of T2K Run periods.}
    \begin{tabular}{|c|c|}
    \hline
    T2K Run & Data Taking Period \\
    \hline
    Run 1 & March 2010 -- June 2010 \\
    Run 2 & November 2010 -- March 2011 \\    
    Run 3 & February 2012 -- June 2012 \\
    Run 4 & October 2012 -- May 2013 \\
    Run 5 & May 2014 -- June 2014 \\
    Run 6 & November 2014 -- June 2015 \\
    Run 7 & February 2016 -- May 2016 \\
    Run 8 & October 2016 -- April 2017 \\
    Run 9 & October 2017 -- May 2018 \\
    Run10 & November 2019 -- February 2020 \\
    Run11 & March 2021 -- April 2021 \\
    \hline 
    \end{tabular}
    \label{tab:t2kruns}
\end{table}

\subsection{ND280}
\begin{figure}[t]
\center{\includegraphics[width=0.6\linewidth]{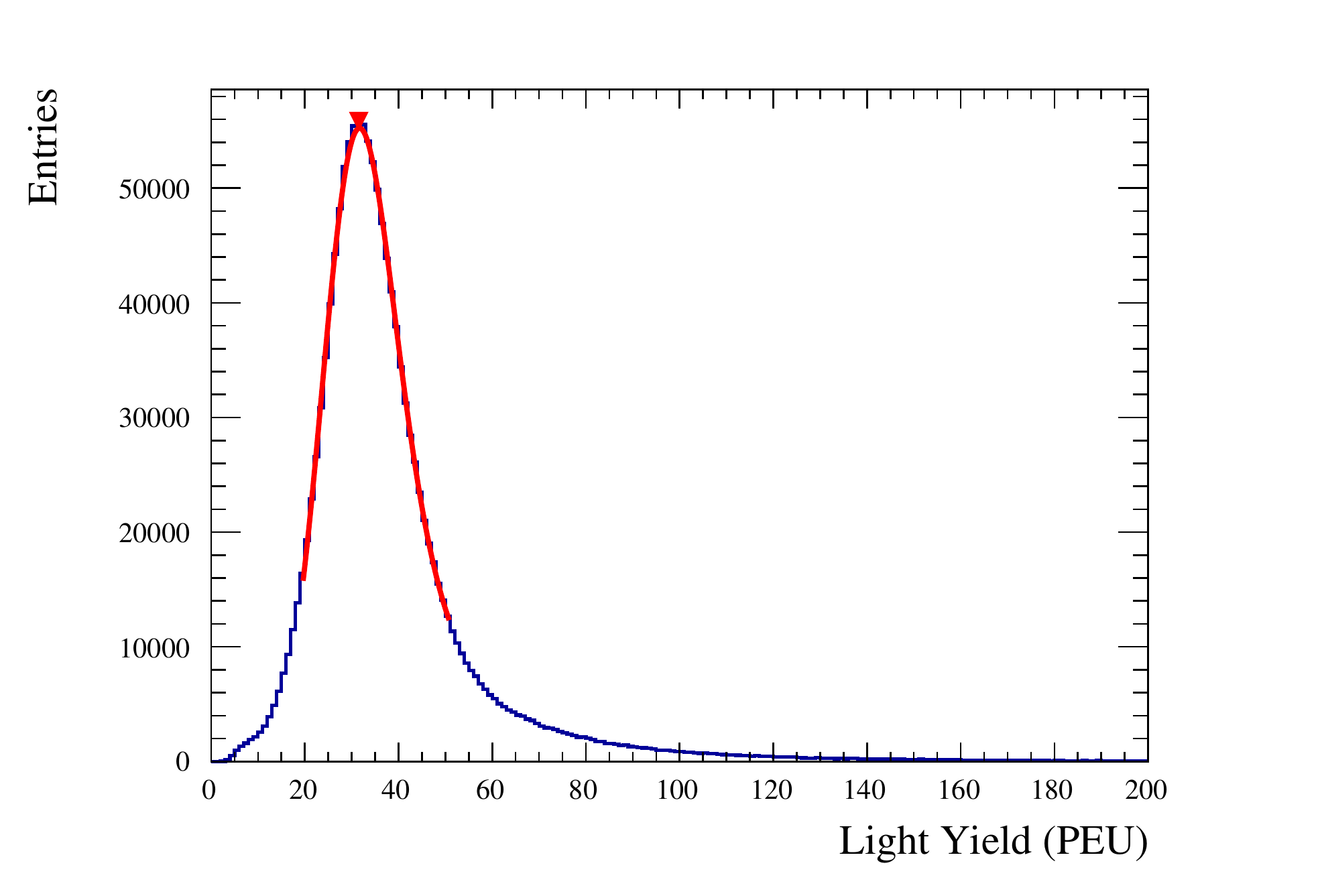}}
    \caption{Example MIP light yield distribution in ECal Barrel X after calibrations and corrections are applied. The MIP most probable value (MPV) in Pixel Equivalent Units (PEU) is extracted from a Landau-Gaussian fit to the distribution. A PEU corresponds to the signal of a single MPPC pixel.}
    \label{fig:peak}
\end{figure}

For all subsystems within ND280, the MPPC response (hits) for MIP-like tracks measured during each T2K Run were combined to create histograms of accumulated charge per unit length. These histograms were then fitted with the convolution of a Gaussian distribution and a Landau distribution \cite{langau}, see figure~\ref{fig:peak}. This distribution models the typical energy loss of high-energy particles in matter, along with a Gaussian term to account for detector smearing effects.

The MIP light yield is taken to be the most probable value (MPV) of the Landau-Gaussian fit function. Calibrations designed to maintain the light yield over time, and therefore account for any ageing of the scintillator, are disabled.

Different MIP track selection criteria were developed for each subsystem, dependent upon the detector geometry and chosen data sample, as described below.

\subsubsection{P\O D} \label{sec:ly_measuremets_p0d}
The P\O D detector uses a sand muon data sample to monitor the scintillator response. This control sample is selected in the following way:
\begin{enumerate}
    \item There is only one 3D track reconstructed within the P\O D during the beam trigger readout window,
    \item This track passes through the first and the last P\O Dule,
    \item Track angle with respect to the beam direction, $\theta$, fulfils the following condition: $\cos\theta \geq 0.8$ (forward going, as measured at the upstream face of the P\O D).
\end{enumerate}
These criteria select a sample of MIPs travelling through the detector, leaving hits with well measured 3D positions. The light yield per unit path length for each individual hit is aggregated for each T2K Run separately for each of the four Super-P\O Dules.

\subsubsection{FGD}
The FGD (as the P\O D) uses a sand muon data sample to monitor the scintillator response. This control sample only includes events where there is just one 3D track reconstructed within each FGD during the beam trigger readout window. The light yield per unit path length for each individual hit is aggregated together for both FGDs for each T2K Run.

\subsubsection{ECal} \label{sec:ly_measuremets_ecal}
During normal detector operation, high statistic samples of cosmic ray muons traversing the ND280 ECals are routinely recorded. These provide an ideal sample by which to monitor and calibrate the response of the detector modules.

\begin{figure}[t]
    \center{\includegraphics[width=0.49\linewidth]{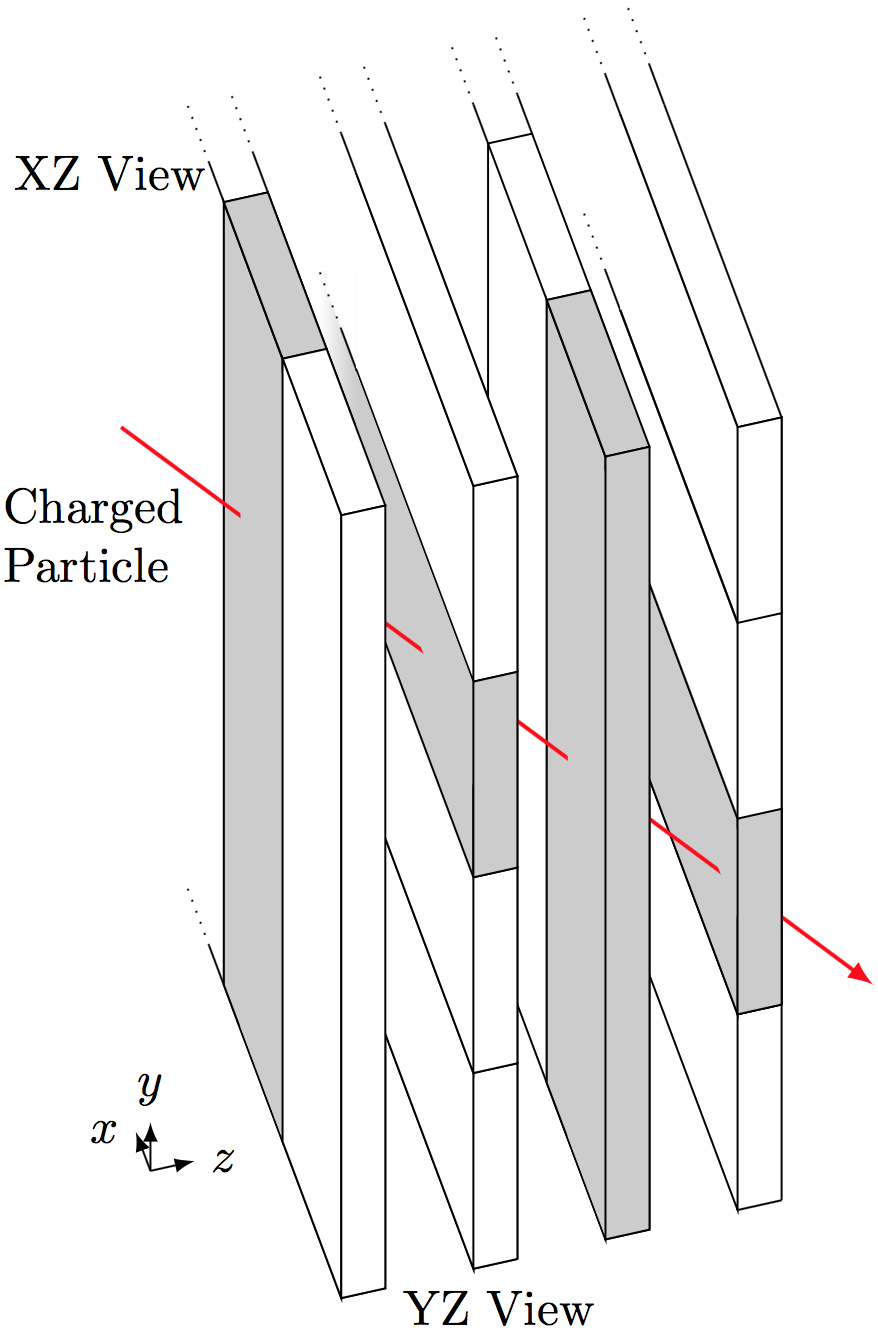}}
    \caption{Schematic view of the passage of a MIP through the Downstream ECal module.}
    \label{fig:ecalrecon}
\end{figure}

The cosmic ray trigger requires the coincidence of hits to occur within two outer regions on opposite sides of the ND280 detector, outside of the time window used for neutrino beam triggers. These hits can occur within the SMRD, Downstream ECal and most upstream Super-P\O Dule and indicate that a cosmic ray has traversed ND280. The MIP tracks are then individually reconstructed in 3D using a linear fitting algorithm, with the hits required to have recorded a valid charge and have adjacent hits in each 2D view, see figure~\ref{fig:ecalrecon}. The calibrated light yield on each bar is then obtained and can be normalised to account for the angle of incidence of the MIP with respect to the scintillator bar, and optionally the attenuation of the scintillation light as it propagates through the WLS fibre to the MPPC. The attenuation correction normalises the response of interactions at any position along the bar to the response observed at 1~m from the MPPC.

The measured light yield of individual MIP interactions are aggregated for each month or T2K Run separately for the different bar lengths described in section \ref{sec:nd280ecal}. The analysis presented here uses a random sampling of 5\% of all ND280 subruns (the data recorded during $\sim$20 minutes of nominal ND280 operation) from each T2K Run to give excellent statistical coverage over all periods of interest. The 3D reconstruction of the MIP tracks also allows for the hits to be aggregated at different positions along the length of the bars, and when the attenuation length correction is disabled allows the light yield to be measured as a function of distance to the MPPC, which is required for the additional studies described in section \ref{sec:scintandfibreageing}.

\subsubsection{SMRD}

For the SMRD both beam and cosmic data samples can be used. However, in most ND280 cosmic trigger configurations the SMRD is not uniformly sampled, leaving some regions statistically limited, unlike in the case of the beam mode triggers. Moreover, only a fraction of the recorded cosmic data sample gets processed. Hence the current study was performed using the beam data sample. The track selection requires:
\begin{enumerate}
    \item The highest momentum track reconstructed within the beam trigger readout window has an interaction vertex within the SMRD fiducial volume,
    \item The track crosses at least one TPC,
    \item The track particle identification hypothesis is consistent with being muon-like.
\end{enumerate}

The light yield per unit path length for each individual SMRD hit is aggregated together for each T2K Run.

\subsection{INGRID}
INGRID uses a high statistics sample of cosmic ray muons to measure the MIP response of the scintillator bars. The recording of cosmic ray muons is triggered when hits are observed near-simultaneously in four scintillator planes of an INGRID module, outside of the neutrino beam trigger timing window. Channels with more than 2.5~PEU (Pixel Equivalent Units) are defined as hits, and 3D track reconstruction from the hits allows for the recorded MIP response to be corrected for the particle's trajectory through the module.

\begin{figure}[t]
\centering
\begin{subfigure}{0.48\textwidth}
    \includegraphics[width=\textwidth]{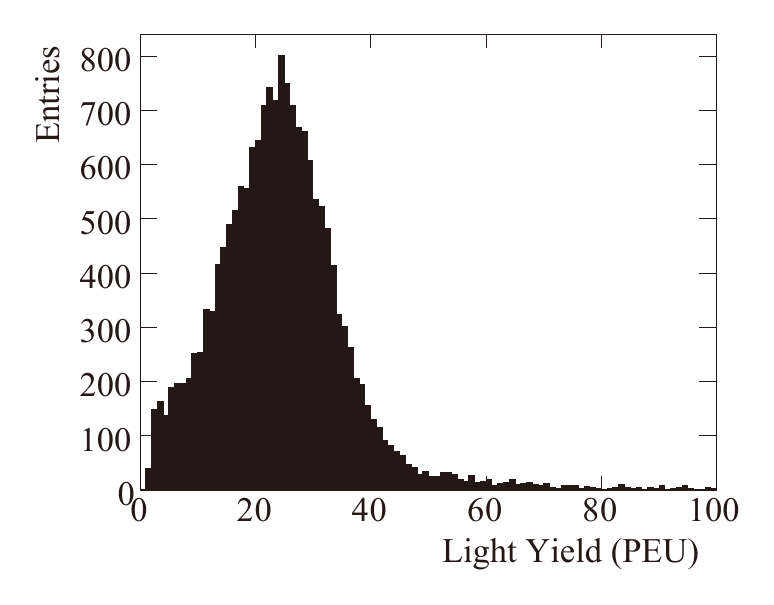}
    \caption{Light yield response of a single channel.}
    \label{fig:ingridMIPChannel}
\end{subfigure}
~
\begin{subfigure}{0.48\textwidth}
    \includegraphics[width=\textwidth]{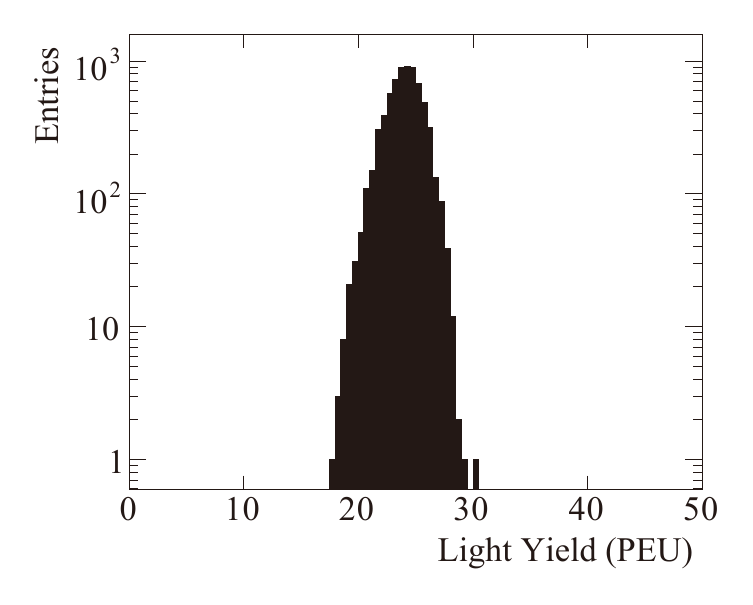}
    \caption{Mean light yield for all channels.}
    \label{fig:MIPMeanOfMeans}
\end{subfigure}
\caption{The typical response of a single readout channel (a) and aggregated mean light yield of all channels (b) for INGRID.}
\label{fig:ingridMIP}
\end{figure}

The INGRID working group has independently assessed the scintillator ageing of the INGRID detector using a different, but equally valid method. Unlike the Landau-Gaussian fitting method employed by the ND280 subsystems, during each J-PARC Main Ring Run (the period between each exchange of the H$^-$ ion source, typically one month) the MIP response distribution of each INGRID readout channel is aggregated and the mean response is extracted, see figure~\ref{fig:ingridMIPChannel}. The mean response of all channels are then aggregated, see figure~\ref{fig:MIPMeanOfMeans}, and the mean of that distribution, the mean-of-means (MOM), is tracked in time to monitor the annual decline in light yield.

\subsection{Light Yield Stability Uncertainty}

During each data aggregation period (time bin) the measured light yield will vary with time due to changes in the stability of the MPPC response, for example due to overvoltage fluctuations caused by changes in ambient temperature. Such fluctuations affect not only the gain but also the photon detection efficiency. Every effort is made to measure and calibrate out these variations, however this is an imperfect process. Therefore, each subsystem attempts to measure the inherent variation in light yield response within each time bin, and then includes that variation as a systematic uncertainty on the ND280 subsystem MIP MPV or INGRID MOM.

This light yield stability uncertainty is combined in quadrature with the uncertainty of the ND280 MIP MPV from the standard Landau-Gaussian convolution fit or INGRID MOM.

\subsubsection{ND280}
For all ND280 subsystems, to assess the light yield stability within each time bin, the contributing data samples were split into shorter (reduced) time periods. Within each reduced time period, the MIP response was fitted with the Landau-Gaussian convolution, and the MIP MPV extracted. For each time bin, the standard deviation of the MIP MPVs for the contributing reduced time periods was calculated and taken as the light yield stability uncertainty.

Due to the variation in event rates for the samples used in the MIP MPV estimation for the different ND280 subsystems, the length of the reduced time periods varies between the subsystems to ensure a good balance between temporal granularity and obtaining sufficient statistics to perform an accurate Landau-Gaussian fit. For the ECal, the high statistics of the cosmic ray sample allows data to be aggregated into periods of $\sim$20~minute duration (the period of one ND280 subrun), however for the FGD and P\O{}D, the slower event rate of sand muon data means the data was aggregated into periods of one-month and two-weeks, respectively. For the SMRD, the T2K Runs with the highest statistics were studied and the data was aggregated into one week periods. The largest standard deviation, among the T2K Runs, was then taken as the error to be conservatively applied to all SMRD data points. The range of uncertainties, and modal uncertainty, across all data periods are shown in table~\ref{tab:lystaberror}. Most uncertainty values lie close to modal value, with a few exceptions which push out the maximum range to higher values.
\begin{table}[t]
    \footnotesize
    \centering
    \caption{Absolute range and modal light yield stability uncertainties in PEU for each subsystem. Also shown are the range and modal uncertainties as a percentage of the recorded MPV in each time bin.}
    \begin{tabular}{|c||c|c||c|c|}
    \hline
    Subsystem & \multicolumn{2}{c||}{Uncertainty Range} & \multicolumn{2}{c|}{Modal Uncertainty} \\
     & Absolute Value (PEU) & \% of MPV & Absolute Value (PEU) & \% of MPV \\
    \hline
    P\O D & 0.02--0.57 & 0.11--2.57 & $\sim$0.20 & $\sim$0.90 \\
    ECal (Single-ended) & 0.07--2.19 & 0.28--8.24 & $\sim$0.15 & $\sim$0.80 \\
    ECal (Double-ended) & 0.05--1.22 & 0.33--7.35 & $\sim$0.10 & $\sim$0.90 \\
    FGD   & 0.10--0.29 & 0.51--1.33 & $\sim$0.15 & $\sim$0.70 \\
    SMRD  & 0.79--1.33 & 1.4--2.3 & 1.33 & 2.3\\
    \hdashline
    INGRID & 0.07--0.74 & 0.30--3.28 & $\sim$0.30 & $\sim$1.50 \\
    \hline 
    \end{tabular}
    \label{tab:lystaberror}
\end{table}

\subsubsection{INGRID}
INGRID takes a similar approach to the ND280 subsystems, aggregating the cosmic ray data over 3 day periods and extracting the standard deviation in the MOM extracted from those reduced periods. This provides uncertainties of 0.3-3.3\% in each time bin. \newline

\section{Annual Light Yield Reduction} \label{sec:ageing}

\begin{figure}[pht]
\center
\begin{subfigure}{0.49\textwidth}
    \center{\includegraphics[width=\textwidth]{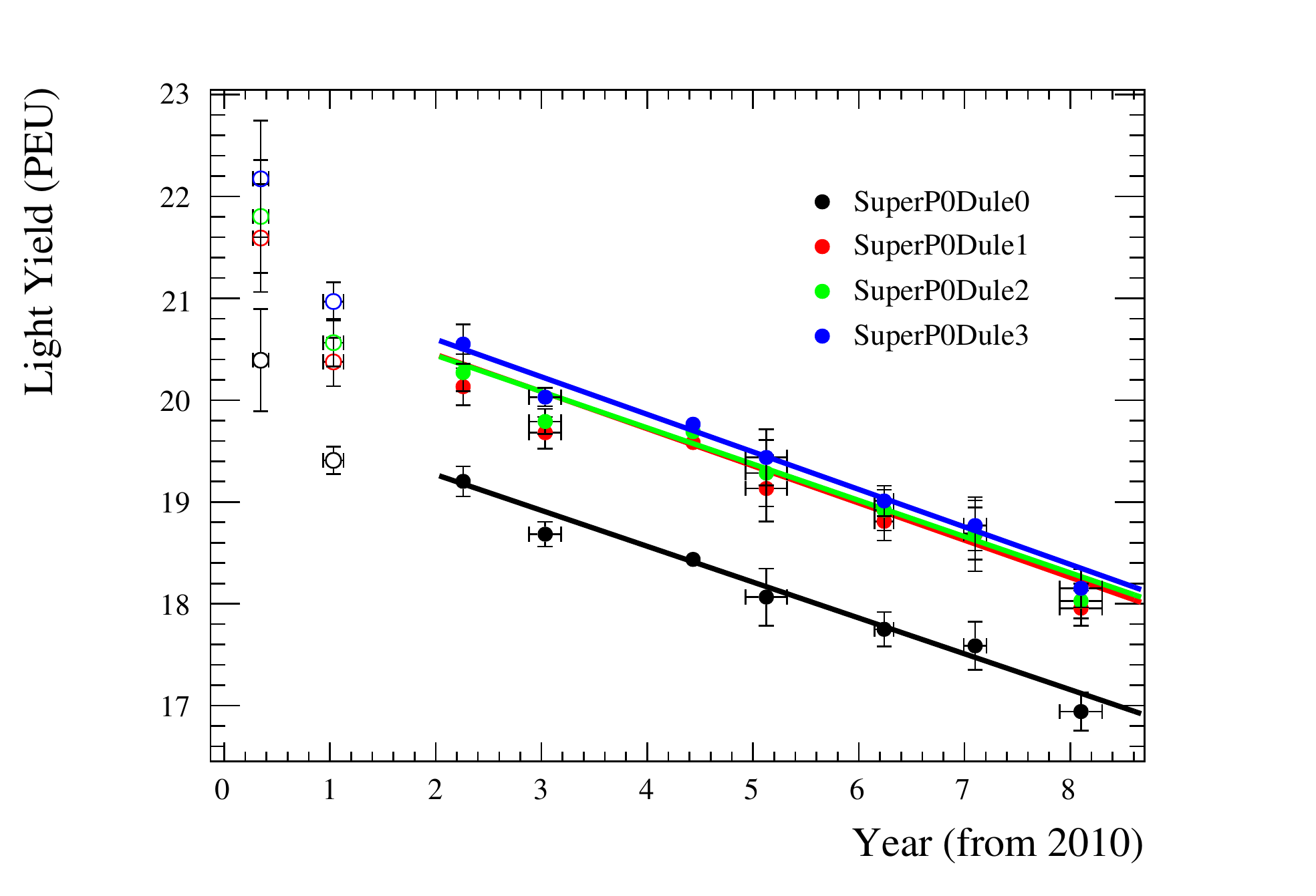}}
    \caption{P\O D.}
    \label{fig:p0d_age}
\end{subfigure}
\hfill
\begin{subfigure}{0.49\textwidth}
    \center{\includegraphics[width=\textwidth]{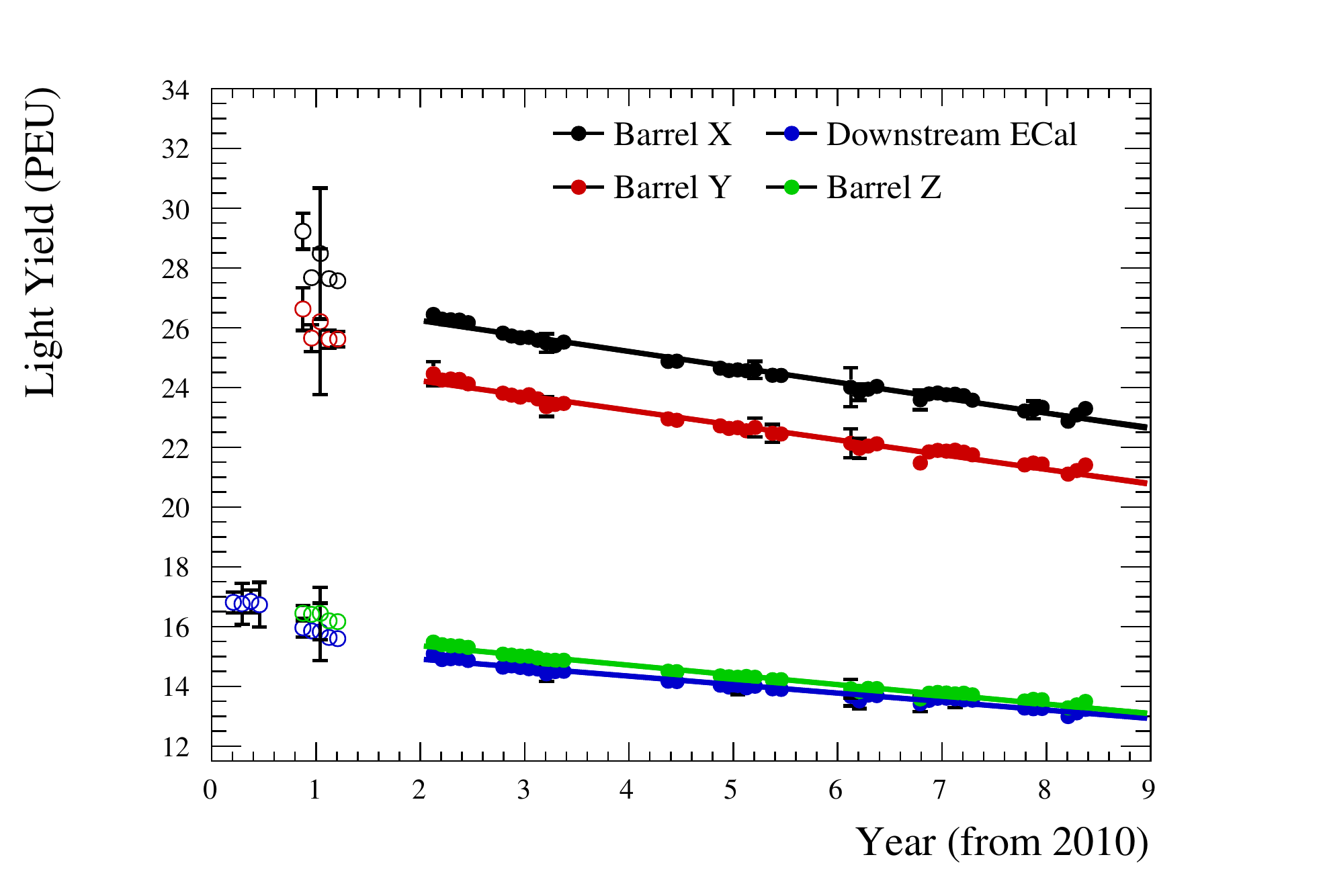}}
    \caption{ECal.}
    \label{fig:ecal_age}
\end{subfigure}
\vfill
\begin{subfigure}{0.49\textwidth}
    \center{\includegraphics[width=\textwidth]{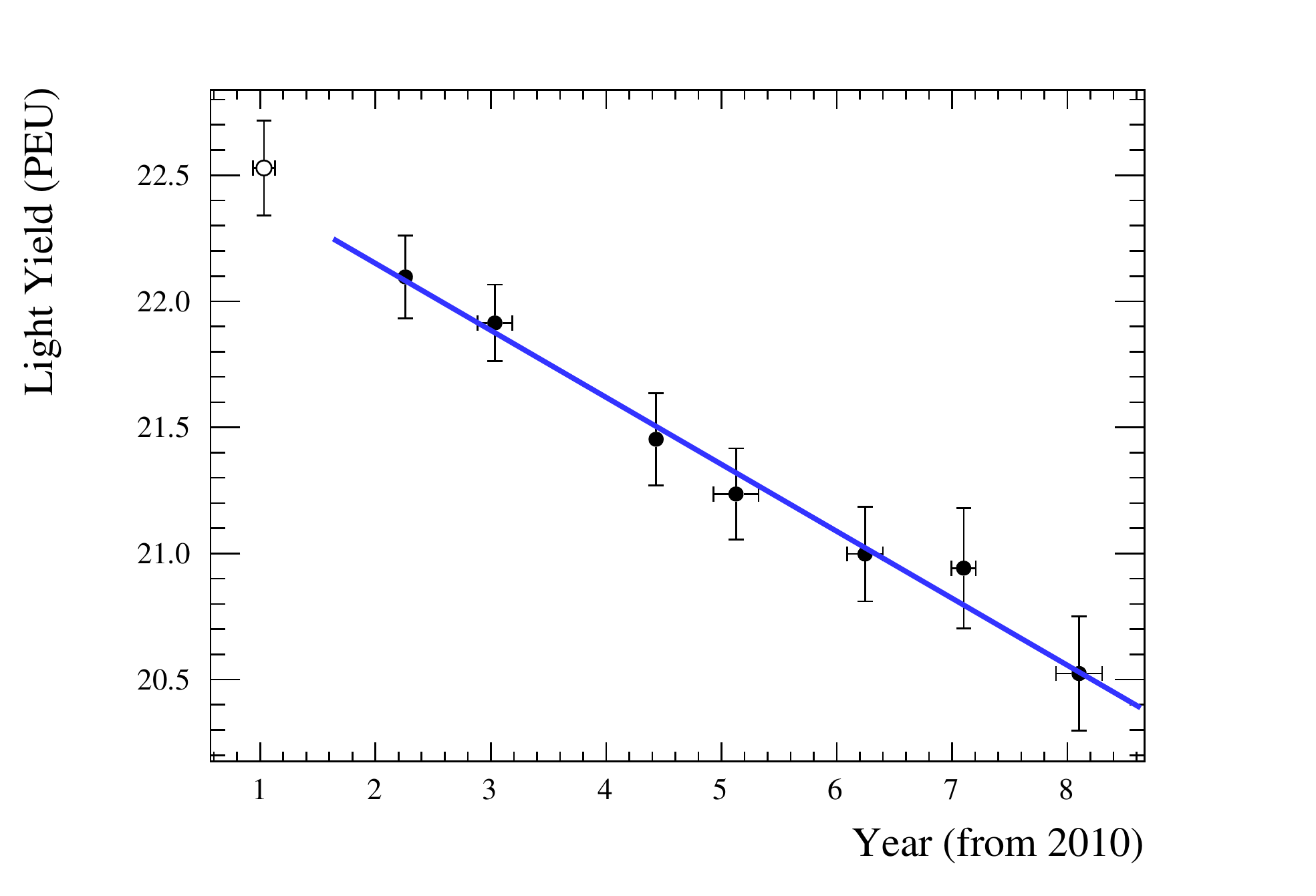}}
    \caption{FGD.}
    \label{fig:fgd_age}
\end{subfigure}
\hfill
\begin{subfigure}{0.49\textwidth}
   \center{\includegraphics[width=\textwidth]{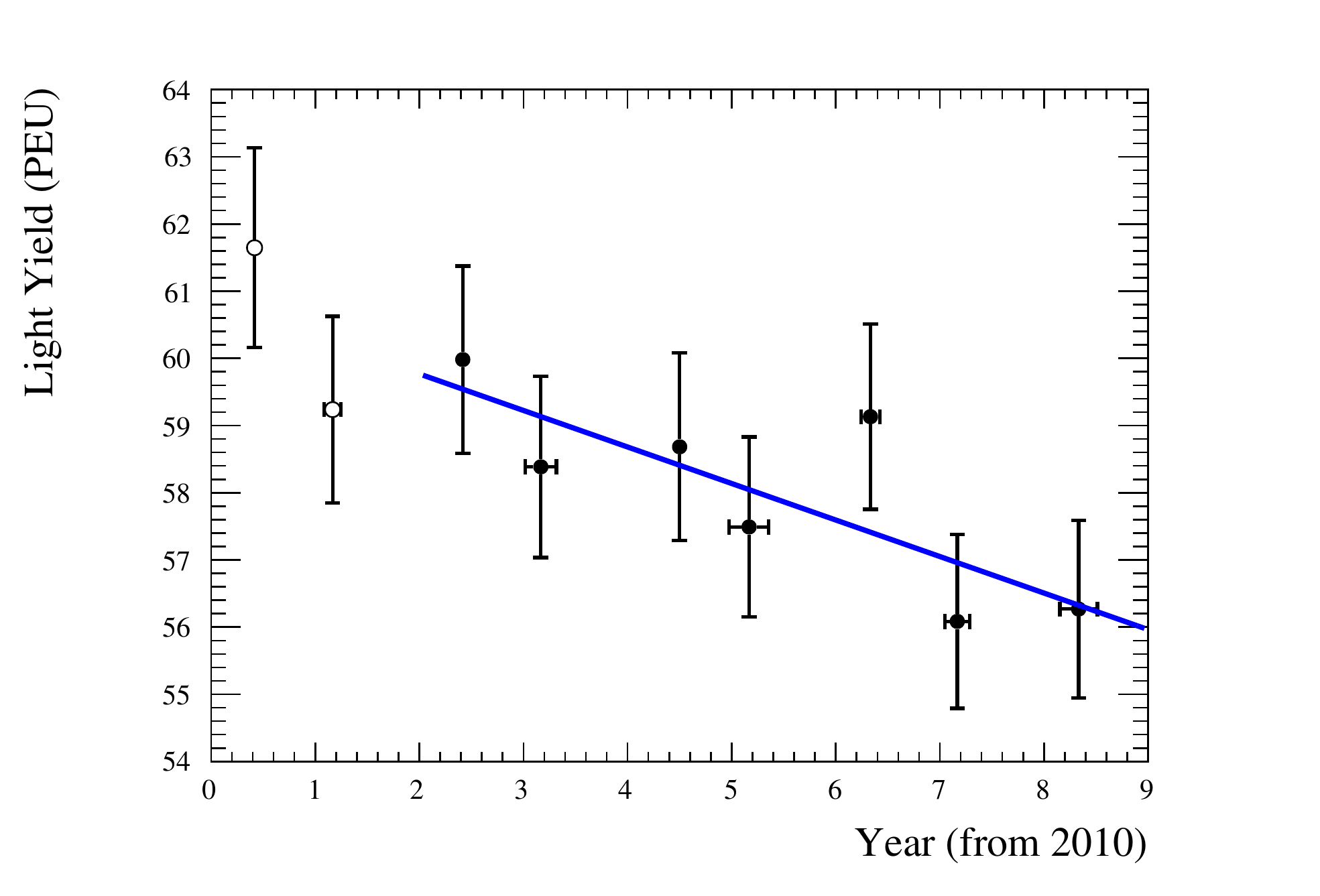}}
    \caption{SMRD.}
    \label{fig:smrd_age}
\end{subfigure}
\vfill
\begin{subfigure}{0.49\textwidth}
    \includegraphics[width=\textwidth]{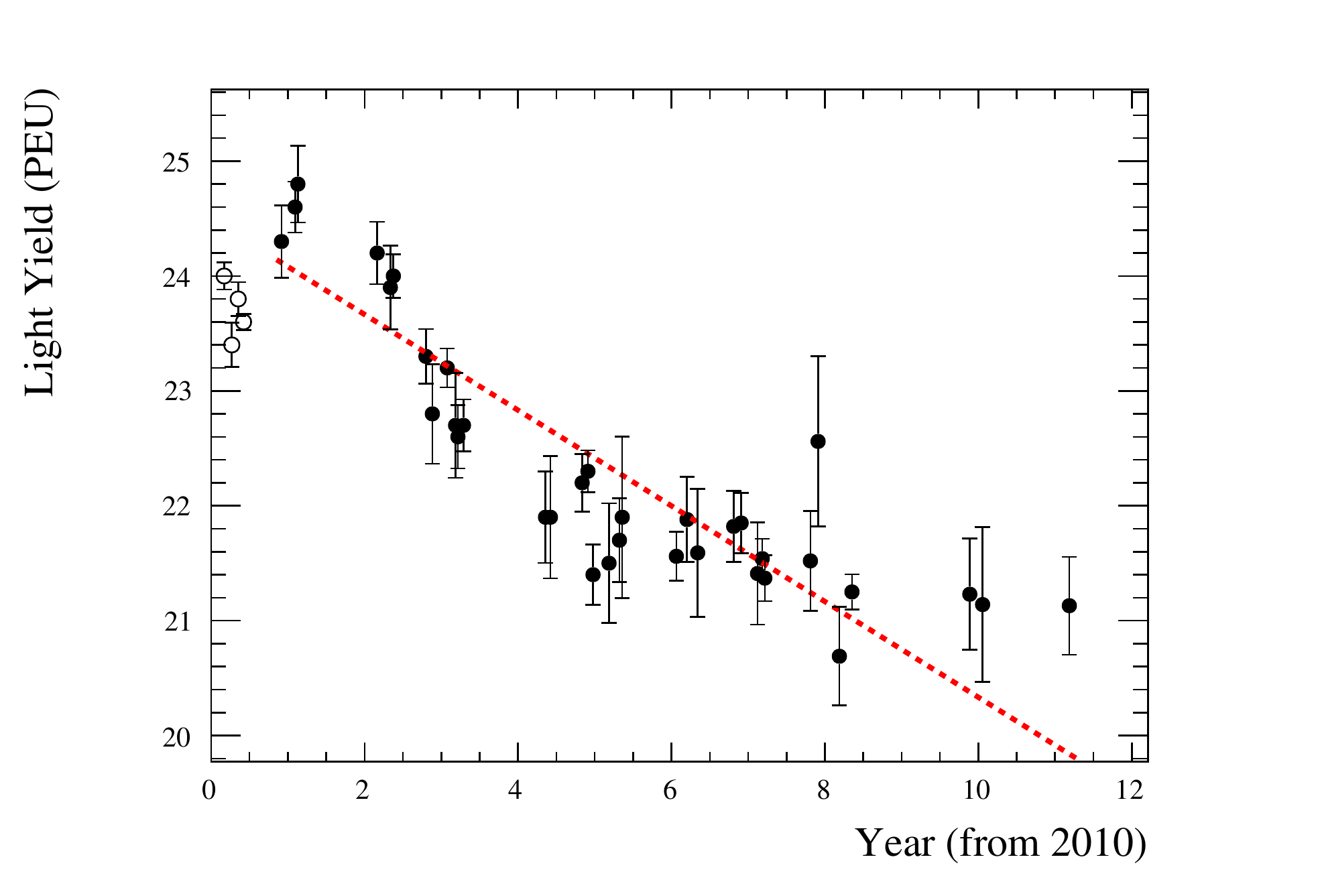}
    \caption{INGRID.}
    \label{fig:ingridAgeingLinear}
\end{subfigure}
\caption{Light yield change in each subsystem for T2K Runs 3--9 (ND280) and Runs 2--11 (INGRID). The x-error bars (time) show the standard deviation in the hit times for each data aggregation period, and the y-error bars (light yield) show the quadratic sum of the light yield stability uncertainty and the uncertainty on the ND280 Landau-Gaussian MIP MPV or INGRID MOM. Hollow data points are excluded from the data fits.}
\label{fig:ageing_plot}
\end{figure}

The distribution of the ND280 MIP MPV or INGRID MOM was extracted for each subsystem during each T2K Run and then fitted with a linear function in order to calculate the overall drop in light yield and annual decrements, see figure~\ref{fig:ageing_plot}.

The data is aggregated by T2K Run for the P\O D, FGD and SMRD, with the time error being the standard deviation in time stamp of all MIP hits during each T2K Run. For the ECal and INGRID, the higher statistic allows for the data to be instead aggregated on a per month basis, or per J-PARC Main Ring Run, respectively. The fit is only applied to the data from January 2012 (December 2010) onwards for the ND280 (INGRID) subsystems as the current calibration procedures and cosmic ray triggering prescale were not finalised until that time.

As described in the previous section, for the P\O D and ECal subsystems (see sections~\ref{sec:ly_measuremets_p0d} and \ref{sec:ly_measuremets_ecal} respectively) the study has been performed for all sub-modules or bar types separately, see figures~\ref{fig:p0d_age} and \ref{fig:ecal_age}.

The reduction in light yield extracted by the described method measures the loss in performance of the whole readout system; the scintillator, WLS fibre, MPPC and their couplings. However it is assumed that the bulk of the light yield reduction can be ascribed to the degradation of the plastic scintillator as this is a well known phenomenon (as described in section \ref{sec:scintageing}), and there has been no obvious degradation in MPPC performance (for example significant drift in overvoltage settings with time), and the stability of the WLS fibre will be addressed in section \ref{sec:scintandfibreageing}.

Without knowledge of the ageing mechanism(s) degrading ND280 and INGRID subsystems it is difficult to know what form the time dependence on the ageing rate should be expected to take. A priori it might be expected that an exponential function would be suitable, and fits of this form are used for projecting the future response of the most important subsystems in section \ref{sec:futureresponse}. However, given the observed data distributions and timescale studied a simple linear fit is found to be appropriate, and are applied in the form:
\begin{equation}
{f\left(t\right)=A-Bt}\,,
\end{equation}
where $A$ is the fitted light yield in PEU at year 0 (2010), $B$ is the gradient of the fit in PEU per year, and $t$ is the year since 2010. The fit parameters are shown in table~\ref{tab:fit_param}.
\begin{table}[t]
    \footnotesize
    \centering
    \caption{Linear fit parameters to P\O D, FGD, SMRD, ECal and INGRID data from figure~\ref{fig:ageing_plot} and the annual percentage reduction, relative to the 2012 fit values. Single-ended readout bars are mirrored on one end.}
    \begin{tabular}{|c|c||c|c|c||c|}
    \hline
    Subsystem & Readout Type & A (PEU) & B (PEU/yr) & $\chi^2/\text{NDF}$ & Annual Reduction (\%) \\
    \hline
        Super-P\O Dule 0 & Single-ended & $19.97\pm0.15$&$0.35\pm0.03$&$4.35/5 = 0.87$ & $1.82\pm0.16$ \\
        Super-P\O Dule 1 & Single-ended & $21.17\pm0.16$&$0.36\pm0.04$&$10.39/5 = 2.08$ & $1.76\pm0.20$ \\
        Super-P\O Dule 2 & Single-ended & $21.15\pm0.17$&$0.36\pm0.03$&$9.14/5 = 1.83$ & $1.76\pm0.15$ \\
        Super-P\O Dule 3 & Single-ended & $21.33\pm0.16$&$0.37\pm0.03$&$5.46/5 = 1.09$ & $1.80\pm0.15$ \\
        FGD  & Single-ended & $22.68\pm0.19$ & $0.27\pm0.04$ & $0.74/5=0.15$ & $1.22\pm0.18$ \\
        SMRD & Double-ended & $60.86\pm1.48$ & $0.54\pm0.26$ &  $2.62/5=0.52$ & $0.90\pm0.44$ \\
        ECal Barrel X & Single-ended & $27.27\pm0.06$ & $0.52\pm0.01$ & $33.09/37=0.89$ & $1.98\pm0.04$ \\
        ECal Barrel Y & Single-ended & $25.21\pm0.08$ & $0.49\pm0.01$ & $31.88/37=0.86$ & $2.02\pm0.04$ \\
        ECal Barrel Z & Double-ended & $16.01\pm0.04$ & $0.33\pm0.01$ & $36.60/37=0.99$ & $2.15\pm0.07$ \\
        Downstream ECal & Double-ended & $15.48\pm0.06$ & $0.28\pm0.01$ & $11.57/37=0.31$ & $1.87\pm0.07$ \\
         \hdashline
         INGRID & Single-ended & $24.50\pm0.11$ & $0.42\pm0.02$ & $89.32/33=2.71$ & $1.78\pm0.08$ \\
        \hline 
    \end{tabular}
    \label{tab:fit_param}
\end{table}

The degradation of the scintillator appears to be reasonably consistent across all subsystems. All show a reduction in light yield within the range $\sim$0.3--0.5~PEU per year, equivalent to an annual light yield reduction of 0.9--2.2\% relative to their 2012 fit values.

The 1\% difference separating the highest and lowest degradation rates between the materially identical FGD (1.2\%) and ECal (Barrel Z 2.2\%), is not surprising given the differences in production dates for the scintillator bars, and the varying environmental conditions they had experienced during their production, transportation, and positioning within ND280. All of the aforementioned factors will have contributed to differences in the temperature, humidity and UV exposure of the bars across their lifetimes, and so impacted upon their respective ageing profiles.

The higher statistics of the ECal allows for a finer assessment of its ageing rate. An initial rapid ageing is observed within the first two years of operation, followed by a near linear reduction beyond 2012, however it is unclear if this is a real effect or just an artefact of the changes in calibration procedure and cosmic ray triggering prescale. The higher ECal light yield obtained by the Barrel X and Y bars is due to the combination of direct and reflected light signals for these single-ended (mirrored) readout channels, compared to direct transmission only for the double-ended readout of the Barrel Z and Downstream bars.

Results from the MINOS experiment, which uses materially identical bars to the FGD, ECal, P\O{}D and INGRID, showed ageing rates of $\sim$2\% per year~\cite{Michael:2008bc} over 3 and 4.5 year periods measured with their near and far detectors respectively, in good agreement with the higher rates we obtain from the ECal and P\O{}D. 

The MINER$\nu$A experiment found a substantially higher rate of ageing for their scintillator bars, equivalent to a $\sim$7.5\% annual reduction in light yield over a 2 year study period~\cite{Aliaga:2013uqz}. It is unclear why MINER$\nu$A measured such a high rate of degradation as their scintillator composition is again identical to that used by MINOS and most T2K subsystems. It might be possible that MINER$\nu$A has sampled an initial rapid ageing of their scintillator, as perhaps indicated in the earliest ECal data points as discussed above, and also anecdotally observed by MINOS~\cite{Michael:2008bc}; and that further study of later data would show a reduced ageing rate in line with those measured by T2K and MINOS. For completeness, if a linear fit is applied only to the currently excluded Downstream ECal data recorded during the 2010--2011 period, an annual light loss rate of $1.33\pm0.29$ PEU per year on an initial light yield of $17.2\pm0.3$ is obtained. This is equivalent to annual reduction in light yield of $7.7\pm1.7$\% which is in excellent agreement with the MINER$\nu$A result.

\section{Projected Future Response} \label{sec:futureresponse}

The P\O{}D subsystem of the ND280 is being decommissioned in 2022 to allow for the upgrade of the ND280 detector \cite{T2K:2019bbb}. The remaining ECal, FGD, SMRD and INGRID subsystems will be retained in their current form, and so an understanding of their future response will become increasingly important as the T2K near detectors continue operating into the T2K-II \cite{Abe:2016tez} and the Hyper-Kamiokande \cite{Abe:2018uyc} eras.

As such the future response of the ECal, FGD and INGRID subsystems has been projected through until 2040. The SMRD is excluded from this study as its initial light yield is substantially higher than for the other subsystems and its rate of degradation is lower. As such the likelihood of the light yield from this subsystem dropping below any reconstruction threshold is not considered to be an issue for the time period considered.

\begin{figure}[t]
\center
\begin{subfigure}{0.49\textwidth}
    \center{\includegraphics[width=\textwidth]{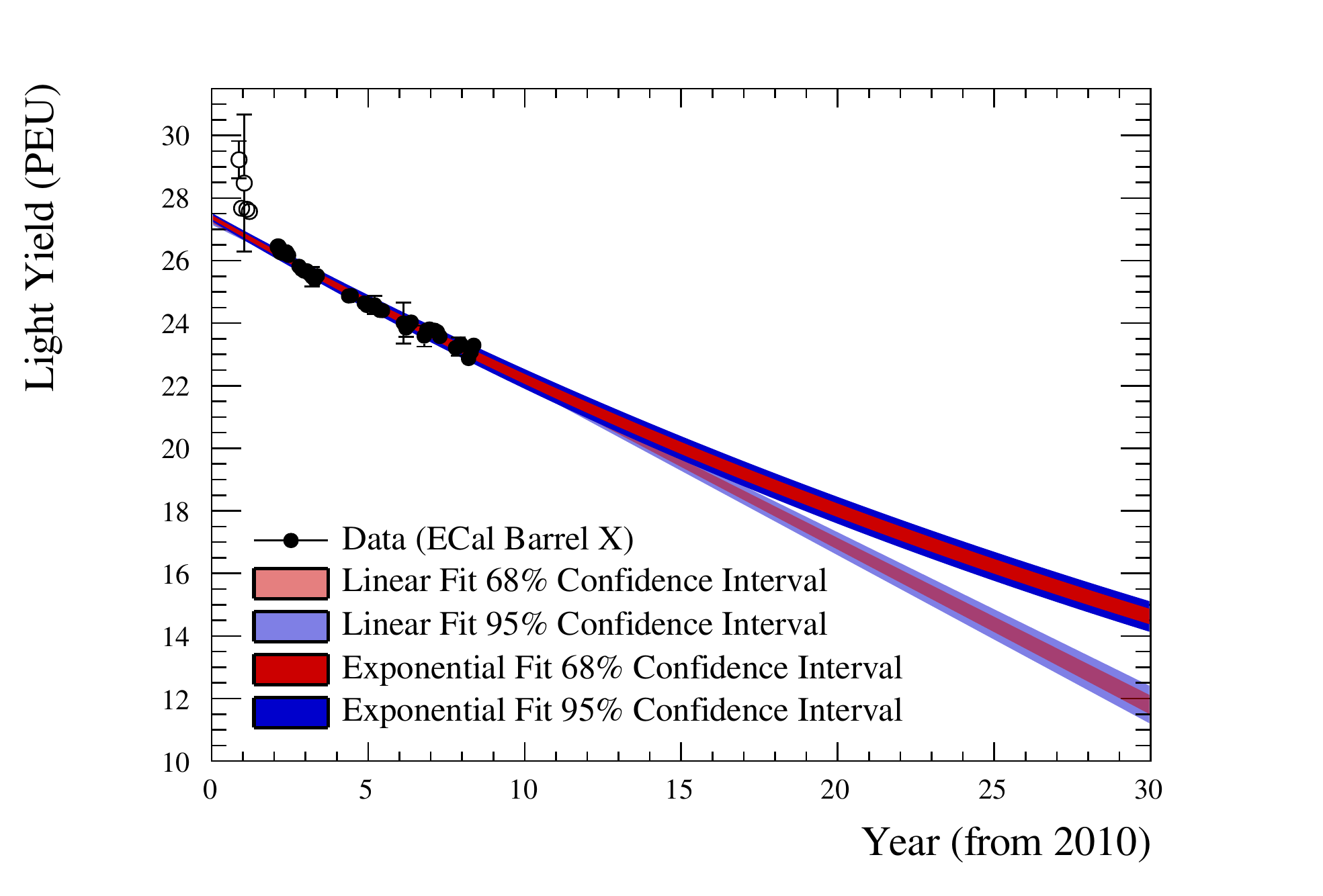}}
    \caption{ECal Barrel X.}
    \label{fig:barrelxprojection}
\end{subfigure}
\hfill
\begin{subfigure}{0.49\textwidth}
    \center{\includegraphics[width=\textwidth]{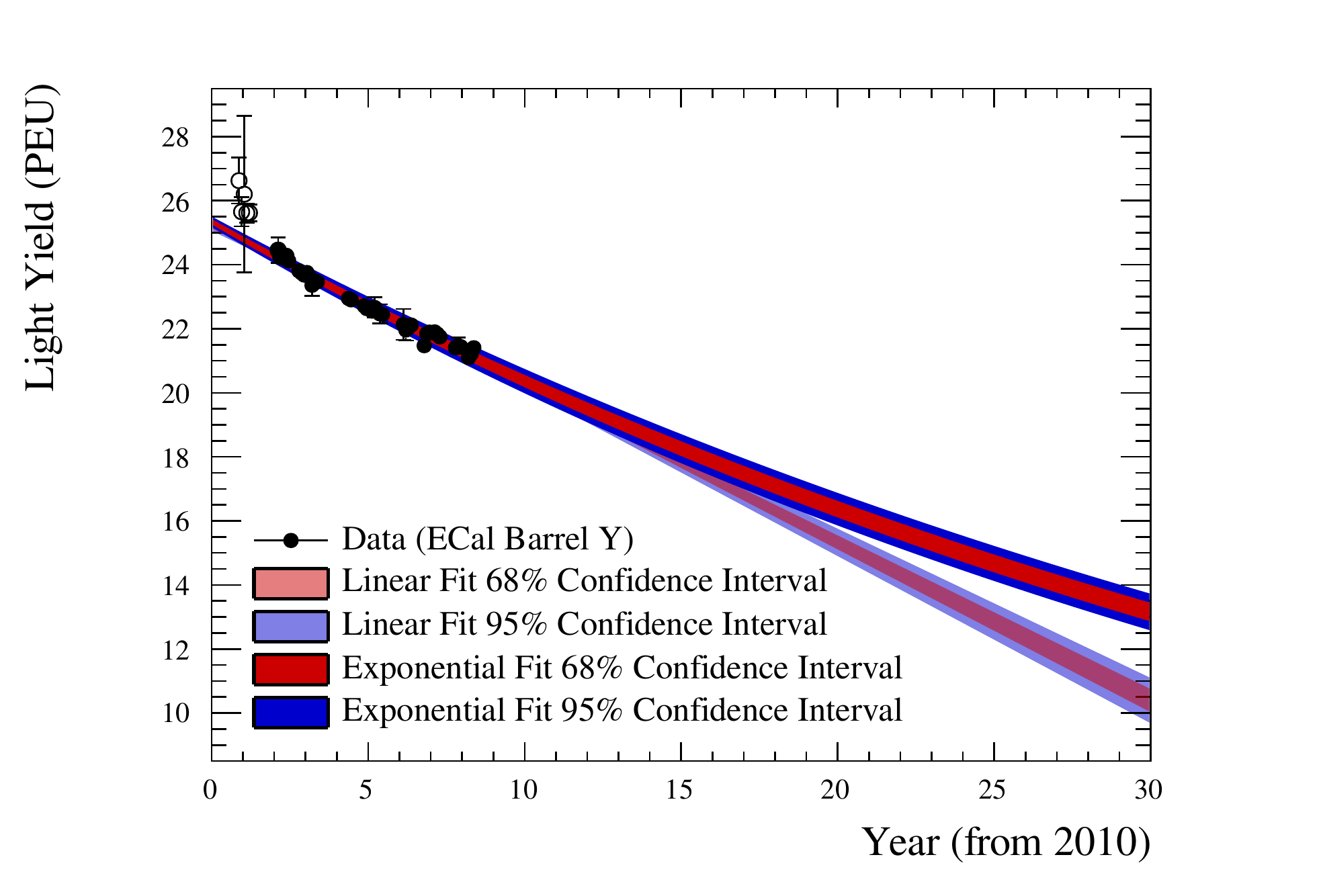}}
    \caption{ECal Barrel Y.}
    \label{fig:barrelyprojection}
\end{subfigure}
\vfill
\begin{subfigure}{0.49\textwidth}
    \center{\includegraphics[width=\textwidth]{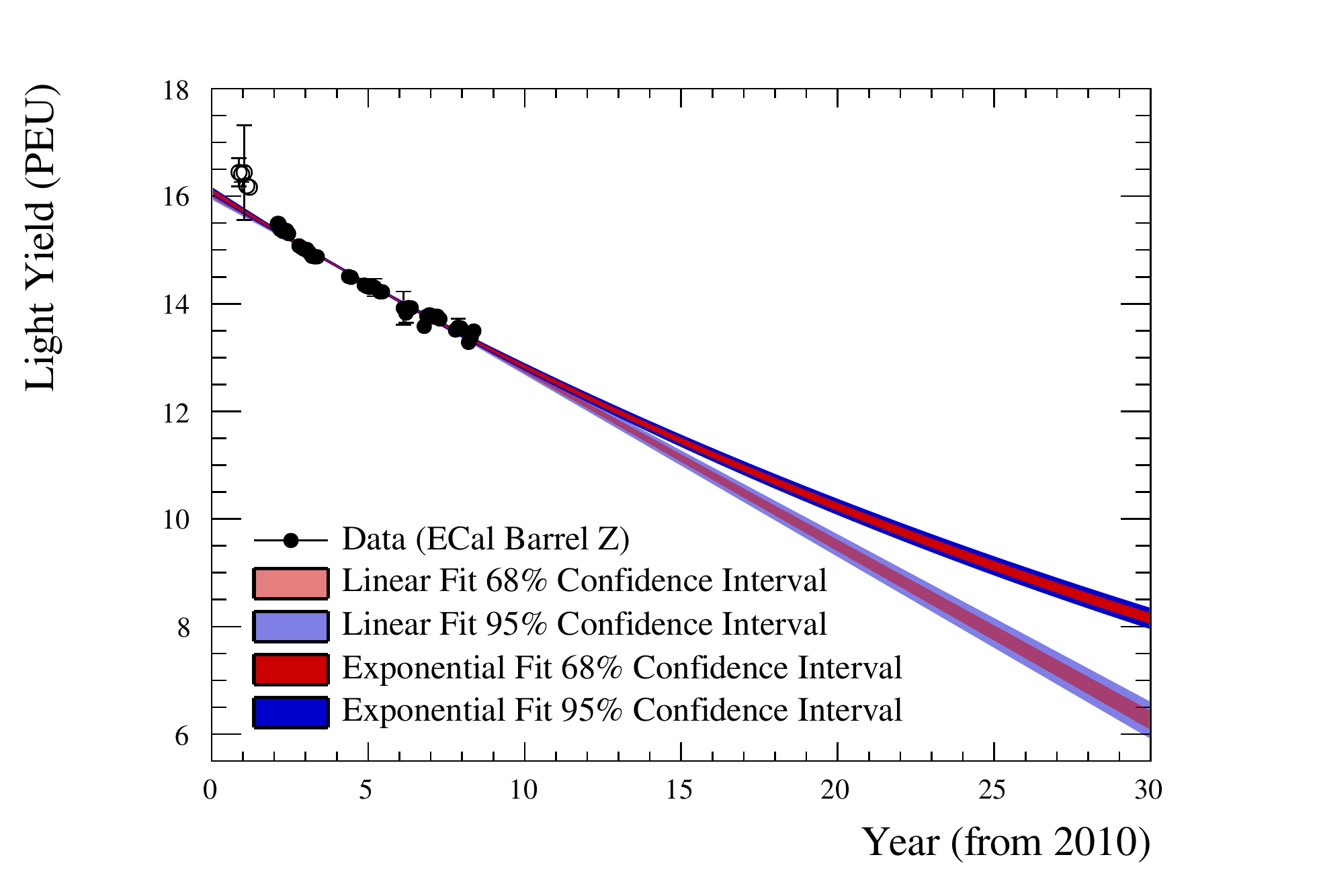}}
    \caption{ECal Barrel Z.}
    \label{fig:barrelzprojection}
\end{subfigure}
\hfill
\begin{subfigure}{0.49\textwidth}
    \center{\includegraphics[width=\textwidth]{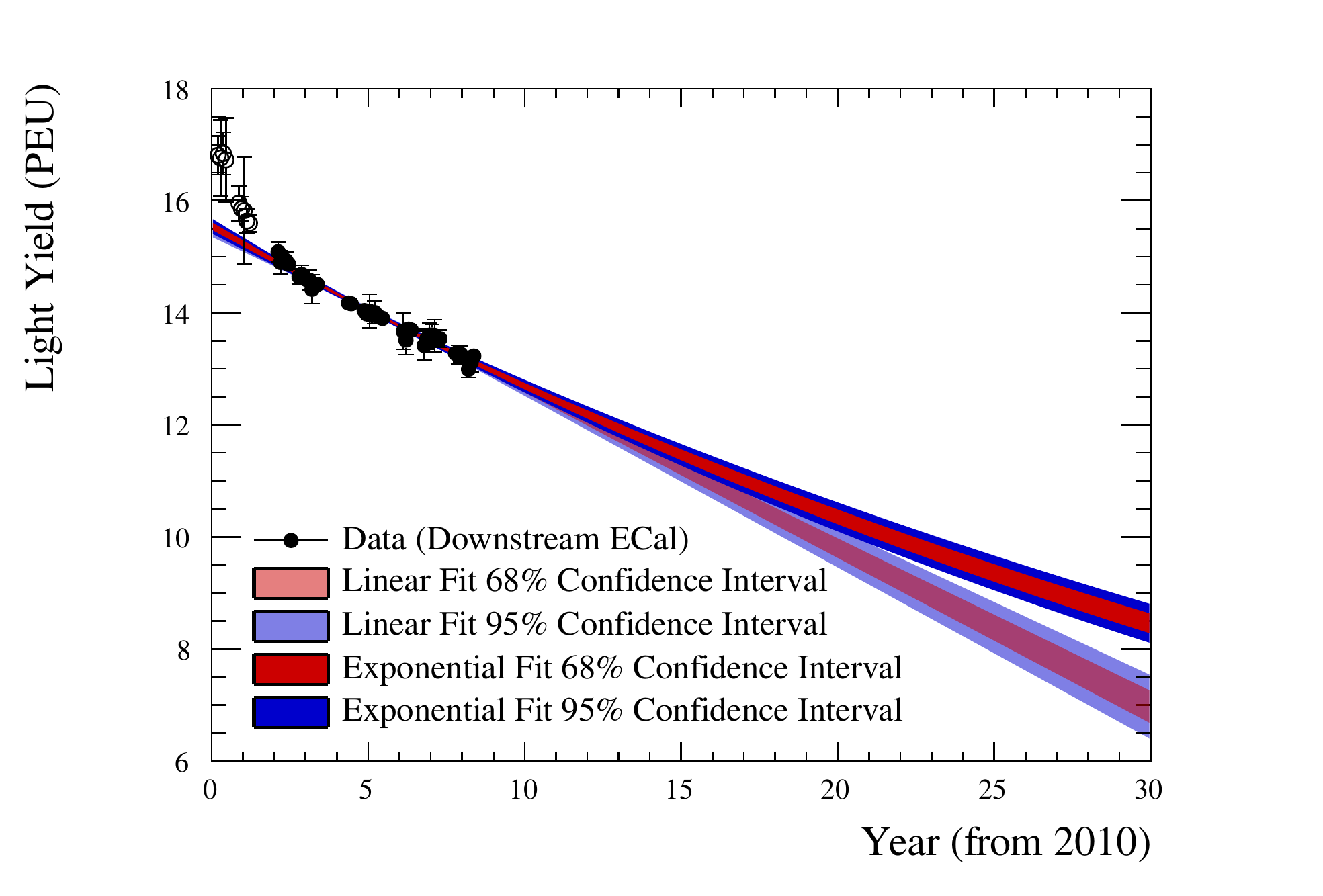}}
    \caption{Downstream ECal.}
    \label{fig:downstreamprojection}
\end{subfigure}
\hfill
\begin{subfigure}{0.49\textwidth}
    \center{\includegraphics[width=\textwidth]{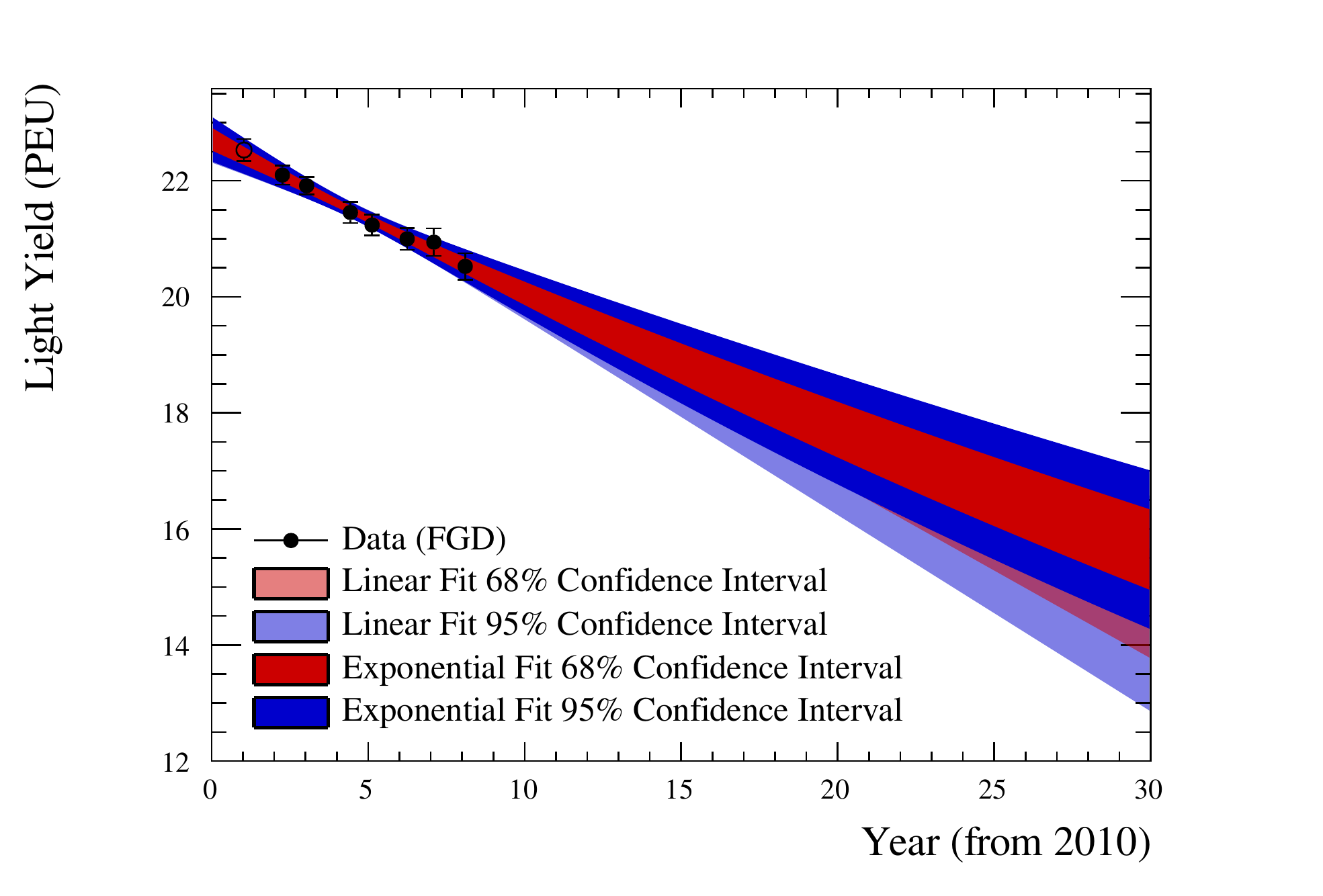}}
    \caption{FGD.}
    \label{fig:fgdprojection}
\end{subfigure}
\hfill
\begin{subfigure}{0.49\textwidth}
    \center{\includegraphics[width=\textwidth]{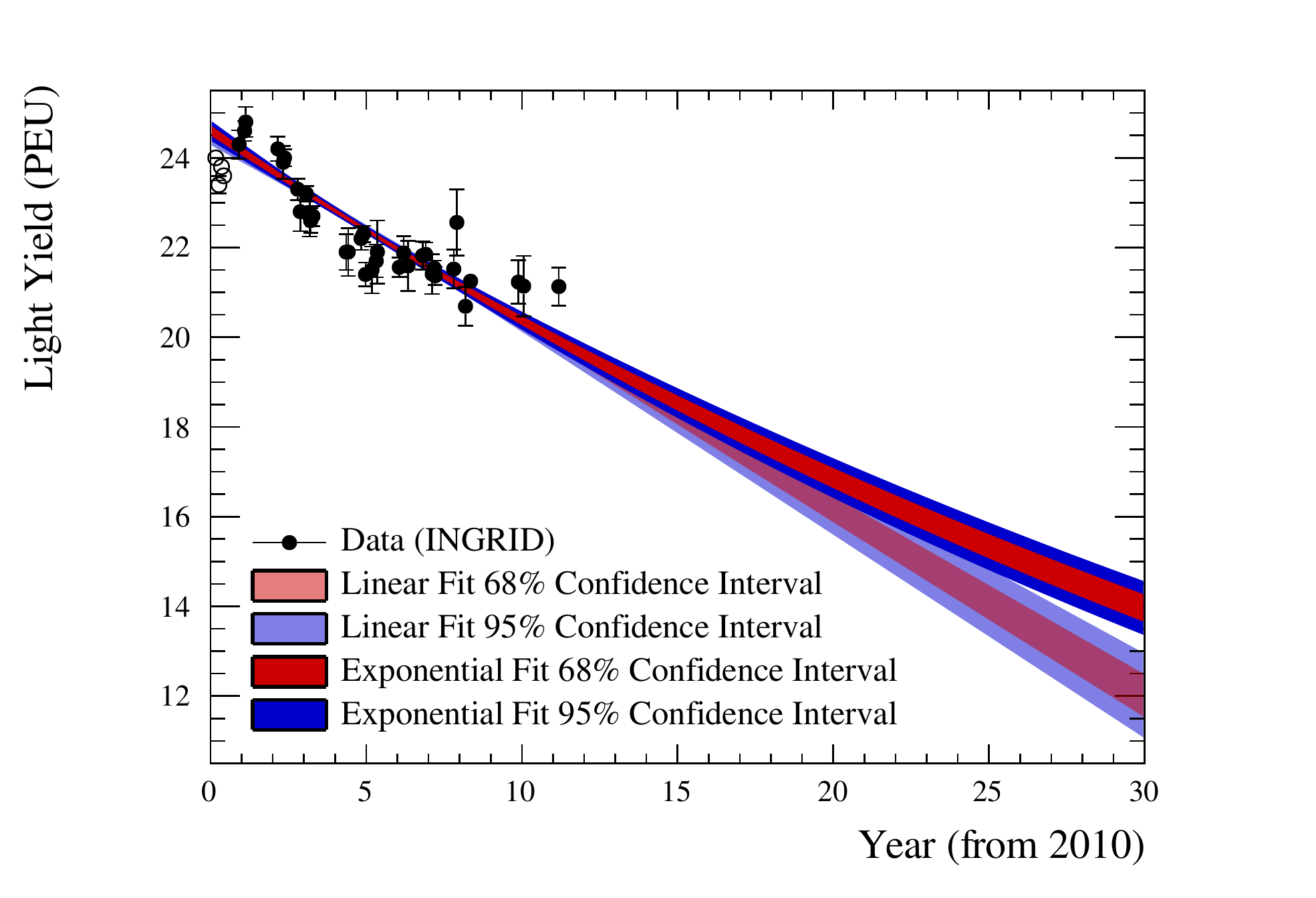}}
    \caption{INGRID.}
    \label{fig:ingridprojection}
\end{subfigure}
\caption{Projected light yield for each ECal bar type, FGD and INGRID, showing the 68\% and 95\% confidence intervals extracted from both the linear and exponential fits to the data. Hollow data points are excluded from the data fits.}
\label{fig:projections}
\end{figure}

Although a linear fit to the data in section~\ref{sec:ageing} results in a reasonable agreement, an exponential fit is better physically motivated. Figure~\ref{fig:projections} shows the projected future response from the earlier linear fits, and from the application of an exponential fit to the ECal and FGD data from 2012, and the INGRID data from 2010, onwards. The exponential fit is of the form:
\begin{equation}
    f\left(t\right) = A\exp\left(\frac{-t}{\tau}\right)\,,
\end{equation}
where $A$ is the fitted light yield in PEU at year 0 (2010), $\tau$ is the time constant of the exponent in years, and $t$ is the year since 2010. The fit parameters are shown in table~\ref{tab:exp_fit}.
\begin{table}[t]
\footnotesize
\centering
    \caption{Exponential fit parameters to ECal, FGD and INGRID data from figure~\ref{fig:projections}.}
    \begin{tabular}{|c|c||c|c|c|}
    \hline
    ECal Bar Type & Readout Type & A (PEU) & $\tau$ (yr) & $\chi^2/\text{NDF}$ \\
    \hline
    Barrel X & Single-ended (mirrored) & $27.39\pm0.07$ & $47.7\pm1.1$ & $27.82/37=0.75$ \\
    Barrel Y & Single-ended (mirrored) & $25.34\pm0.08$ & $45.7\pm1.3$ & $28.27/37=0.76$ \\
    Barrel Z & Double-ended & $16.10\pm0.04$ & $44.1\pm0.9$ & $29.64/37=0.80$ \\
    Downstream & Double-ended & $15.55\pm0.07$ & $49.2\pm2.6$ & $10.28/37=0.28$ \\
    FGD & Single-ended (mirrored) & $22.72\pm0.20$ & $80.3\pm11.1$ & $0.68/5=0.14$ \\
    \hdashline
    INGRID & Single-ended (mirrored) & $24.61\pm0.11$ & $52.9\pm2.4$ & $82.27/33=2.49$ \\
    \hline
    \end{tabular}
    \label{tab:exp_fit}
\end{table}

The time constant $\tau$ is consistent, $\sim$44--49~years, for all ECal bar types, along with the light yield constant $A$ for the pairs of single-ended (mirrored), $\sim$26~PEU, and double-ended, $\sim$16~PEU, readout bars. The resultant $\chi^2/\text{NDF}$ for the exponential fits are marginally reduced by $\sim$0.1--0.2 compared to the corresponding linear fits shown in table~\ref{tab:fit_param}.

The INGRID time constant of $52.9\pm2.4$~years is marginally longer that those recorded by the ECal, and the resultant $\chi^2/\text{NDF}$ for the exponential fit has slightly reduced by 0.22. The FGD records a significantly longer time constant of $80.3\pm11.1$~years and a negligible increase of 0.01 in its $\chi^2/\text{NDF}$ for the exponential fit compared to the linear fit. As with the linear ageing results (see table \ref{tab:fit_param}), some variation in the degradation rates between the different subsystems is expected due to the varying age and environmental exposure profiles of the scintillator bars, although why the FGD should be such an outlier is unclear.

The anticipated ECal response drops by $\sim$50\% or $\sim$60\% over thirty years for all bar types from extrapolations of the exponential and linear fits, respectively. This remains above the minimum charge threshold of 5.5~PEU required for use in the current ECal offline reconstruction algorithms. The value of this threshold has been chosen to avoid discrepancies between data and the current MC simulation at low charges. It should be possible to lower the current charge threshold through more detailed simulation of the detector response, for example including bar non-uniformity and improved MPPC dark-noise rate, and through enhancing the reconstruction algorithms. Without improvement there is a risk that information will be lost for particle interactions which deposit energy at values below the MIP MPV, potentially limiting the physics reach of analyses which utilise the ECals.

The FGD and INGRID subsystems expect their MIP MPV or MOM response to reduce by $\sim$30\% and $\sim$40\%, respectively, over thirty years under the hypothesis of an exponential decline. For both detectors this increases by a further $\sim$5-10\% for a linear decline. For both scenarios this remains far above the hit thresholds of 5.0 and 2.5~PEU, respectively, currently used by the offline reconstruction algorithms for these detectors. If the true rate of ageing were to be higher, such as the $\sim$50--60\% light yield reduction currently projected by the ECal, this would still not be a concern for these subsystems.

\section{Separation of ECal Scintillator and Fibre Degradation} \label{sec:scintandfibreageing}

\begin{figure}[t]
\center
\begin{subfigure}{0.49\textwidth}
    \center{\includegraphics[width=\textwidth]{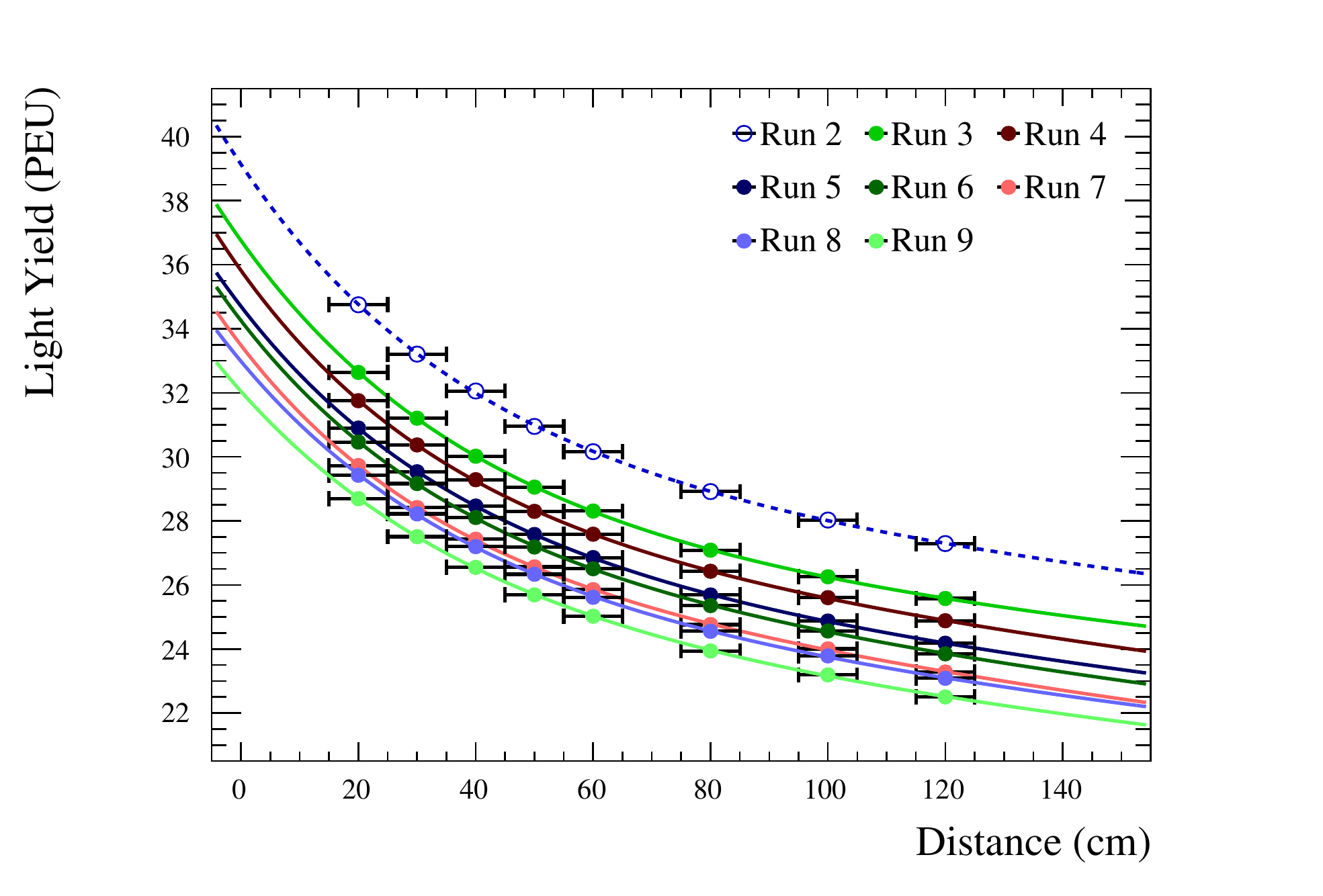}}
    \caption{ECal Barrel X.}
    \label{fig:barrelxdistancempv}
\end{subfigure}
\hfill
\begin{subfigure}{0.49\textwidth}
    \center{\includegraphics[width=\textwidth]{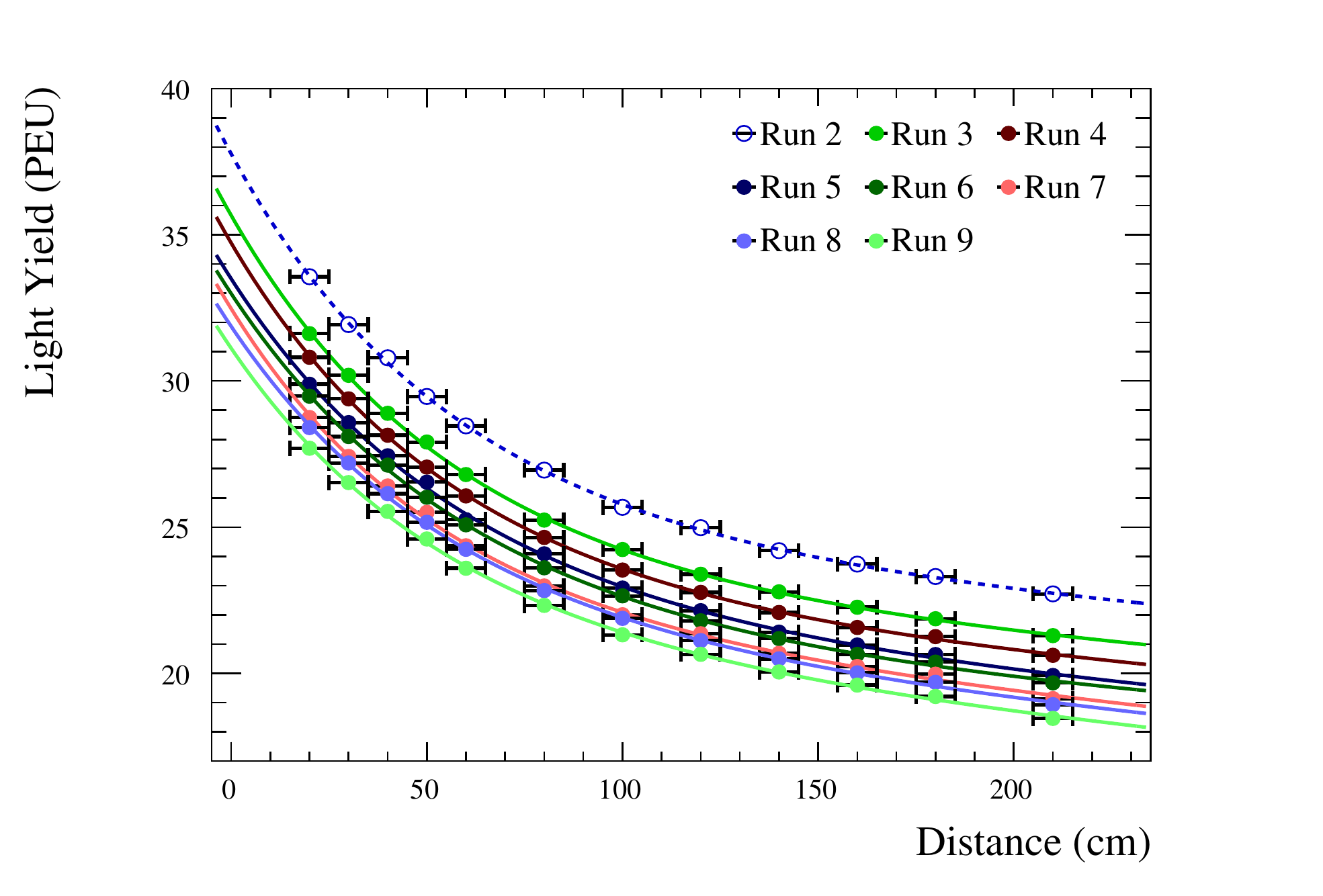}}
    \caption{ECal Barrel Y.}
    \label{fig:barrelydistancempv}
\end{subfigure}
\vfill
\begin{subfigure}{0.49\textwidth}
    \center{\includegraphics[width=\textwidth]{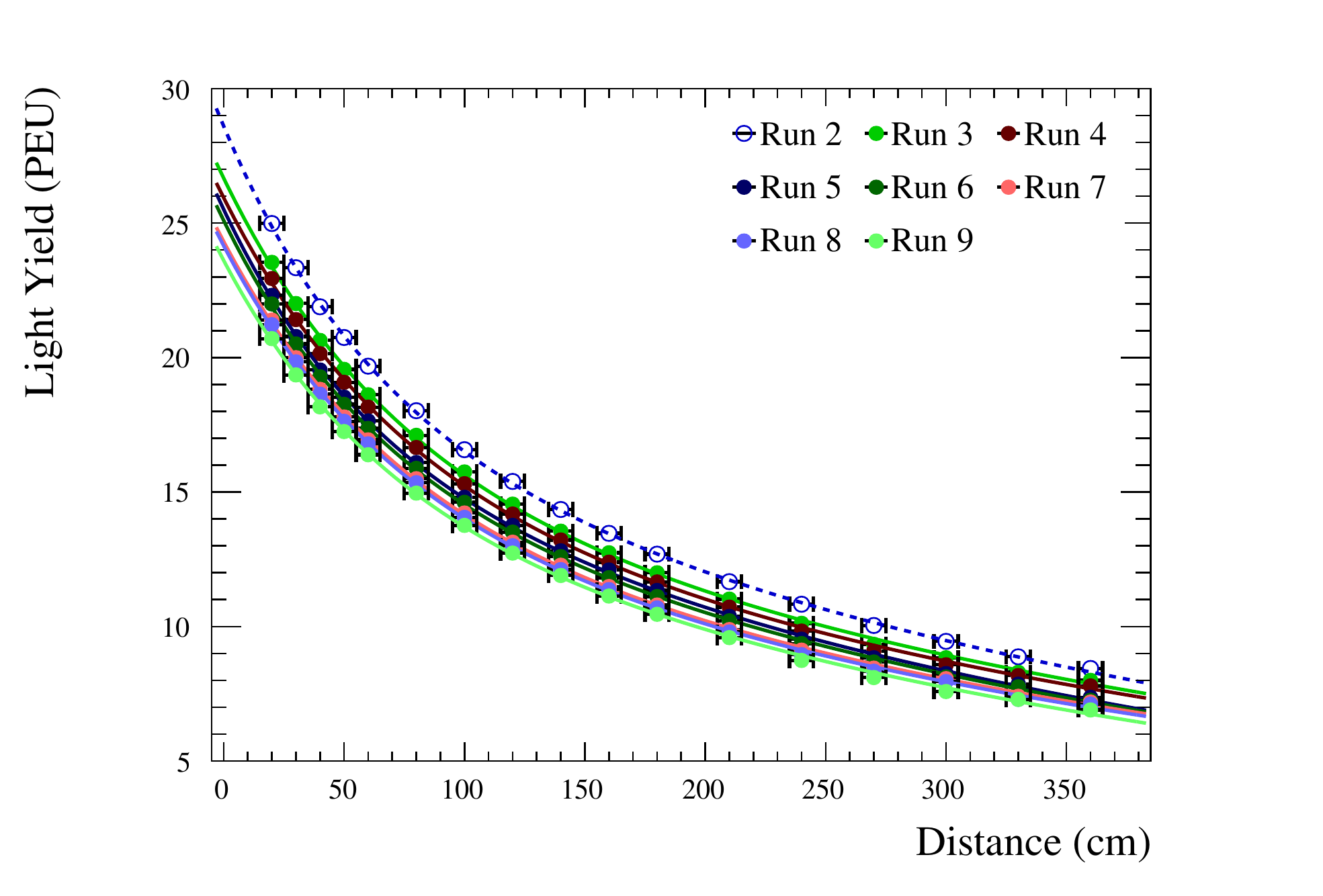}}
    \caption{ECal Barrel Z.}
    \label{fig:barrelzdistancempv}
\end{subfigure}
\hfill
\begin{subfigure}{0.49\textwidth}
    \center{\includegraphics[width=\textwidth]{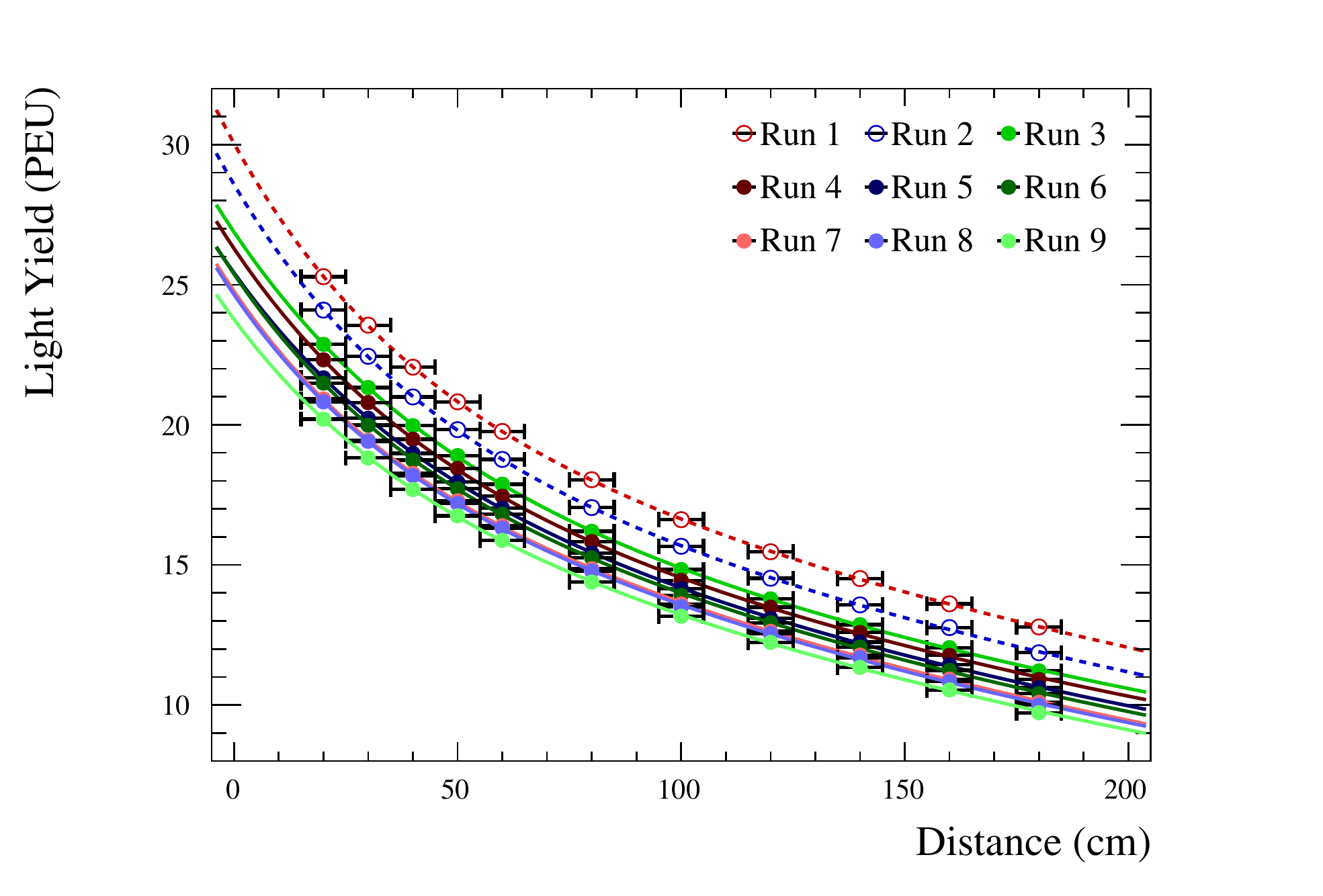}}
    \caption{Downstream ECal.}
    \label{fig:downstreamdistancempv}
\end{subfigure}
\caption{ECal light yield as a function of distance to the MPPC for each T2K Run. The errors on the data points are only the uncertainty on the Landau-Gaussian fit MPV at each distance point, no light yield stability uncertainty is applied. Results of the fits to the hollow data points are excluded from the subsequent data fits.}
\label{fig:ecaldistancempv}
\end{figure}

The results shown in section \ref{sec:ageing} combines the ageing of the scintillator bars with that of the WLS fibres\footnote{Any degradation of the coupling between the fibre and the MPPC, either through loss of transparency of the epoxy or gradual displacement of the fibre also contributes to the results.}, therefore a second approach was applied to separate the two effects within the ECal data. Without applying the attenuation correction, the MIP MPV response is extracted at different distances from the sensor for each bar type during each T2K Run, see figure~\ref{fig:ecaldistancempv}.

The best fit to the data was achieved by applying a double-exponential fit, which accounts for the short and long components of the fibre attenuation, of the form:
\begin{equation} \label{eq:doubleexp}
    f\left(x\right) = S\exp\left(\frac{-x}{\lambda_S}\right)+L\exp\left(\frac{-x}{\lambda_L}\right)\,,
\end{equation}
where $S$ and $L$ are the fitted light yield in PEU at 0~cm from the MPPC for the short and long components of the exponential function, respectively; $\lambda_S$ and $\lambda_L$ are the associated short and long attenuation lengths; and $x$ is the distance from the MPPC in cm.

\subsection{Scintillator Degradation}

\begin{figure}[t]
\centering
    \center{\includegraphics[width=0.49\textwidth]{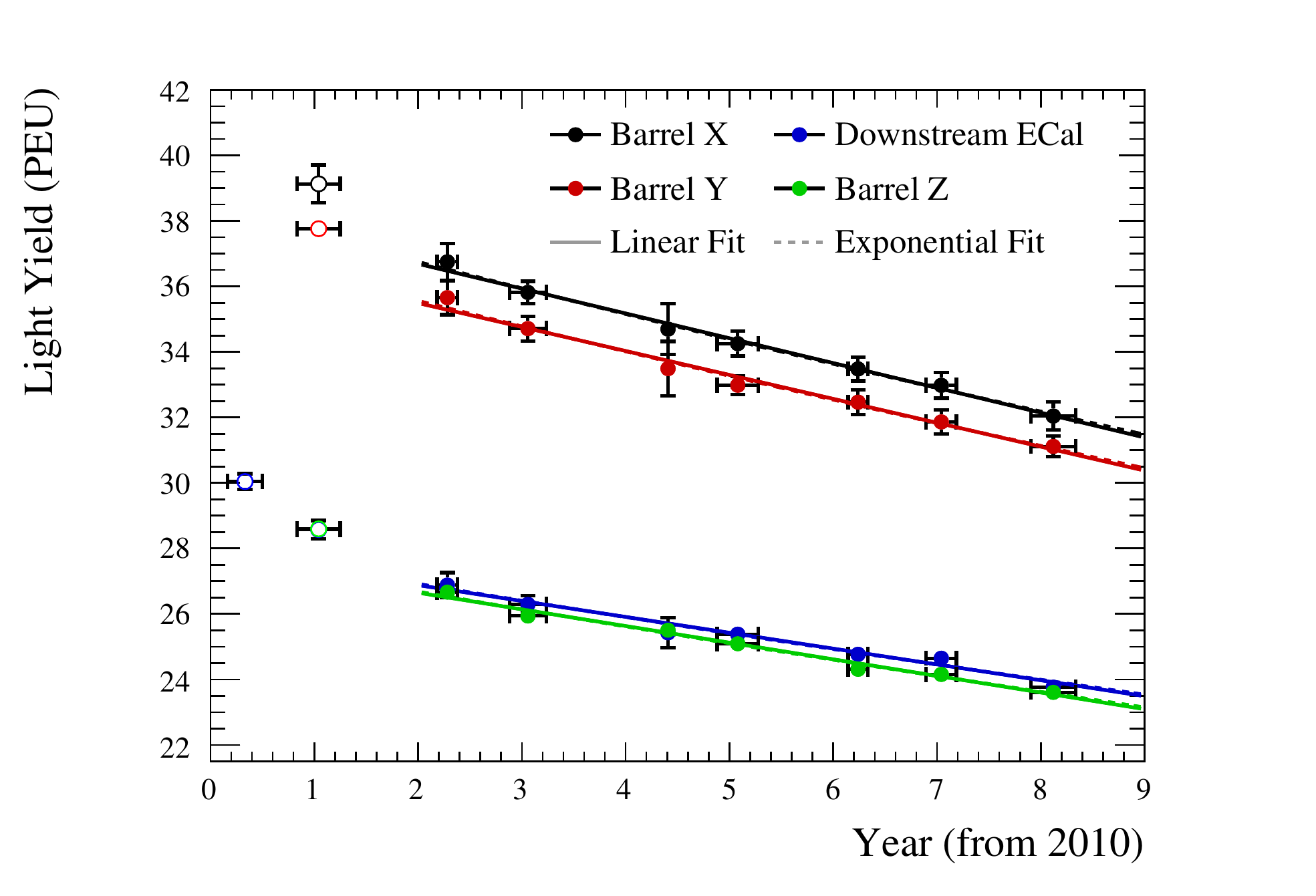}}
\caption{Light yield as evaluated at 0\,cm from the MPPC for each ECal bar type. Hollow data points are excluded from the data fits.}
\label{fig:ecal_distance_ly}
\end{figure}

The parameters of the double-exponential fits can be used to calculate the predicted total MIP light yield at a distance of 0~cm from the MPPC, $S+L$. This should remove the dependence on the propagation of the light down the WLS fibre and the decrease in evaluated light yield will only be dependent on the ageing of the scintillator. The results for this evaluation are shown in figure~\ref{fig:ecal_distance_ly} with both a linear and exponential fit applied to the data from 2012.

The linear fit is of the form:
\begin{equation}
{f\left(t\right)=A-Bt}\,,
\end{equation}
where $A$ is the fitted total MIP light yield ($S+L$) in PEU at year 0 (2010), $B$ is the gradient of the fit in PEU per year, and $t$ is the time in years since 2010. The fit parameters are shown in table~\ref{tab:ecal_lightyield_linear_fits}.
\begin{table}[t]
    \footnotesize
    \centering
    \caption{Linear fit parameters to ECal light yield at 0\,cm from the MPPC from figure~\ref{fig:ecal_distance_ly}, and annual percentage reduction in light yield, relative to 2012 fit value. Reference results, in parentheses, from the linear fit in table~\ref{tab:fit_param} are included for comparison}
    \begin{tabular}{|c|c||c|c|c||c|}
    \hline
    ECal Bar Type & Readout Type & A (PEU) & B (PEU/yr) & $\chi^2/\text{NDF}$ & Annual Reduction (Ref.) (\%) \\
    \hline
    Barrel X & Single-ended & $38.21 \pm 0.50$ & $0.76 \pm 0.09$ & $0.45/5 = 0.09$ & $2.07\pm0.25$ \quad ($1.98\pm0.04$) \\
    Barrel Y & Single-ended & $36.94 \pm 0.47$ & $0.73 \pm 0.08$ & $1.40/5 = 0.28$ & $2.06\pm0.23$ \quad ($2.02\pm0.05$) \\
    Barrel Z & Double-ended & $27.65 \pm 0.18$ & $0.50 \pm 0.03$ & $3.66/5 = 0.73$ & $1.88\pm0.11$ \quad ($2.15\pm0.07$) \\
    Downstream & Double-ended & $27.87 \pm 0.32$ & $0.49 \pm 0.05$ & $2.69/5 = 0.54$ & $1.82\pm0.18$ \quad ($1.87\pm0.07$) \\
    \hline 
    \end{tabular}
    \label{tab:ecal_lightyield_linear_fits}
\end{table}

For the single-ended (mirrored) readout bars the reduction in light yield from the scintillator ageing is $\sim$0.75~PEU per year, and for the double-ended readout bars it is $\sim$0.50~PEU per year. This is a reduction of $\sim$2.1\% for the single-ended (mirrored) bars, and $\sim$1.9\% per year for the double-ended bars.

The exponential fit is of the form:
\begin{equation}
    f\left(t\right) = A\exp\left(\frac{-t}{\tau}\right)\,,
\end{equation}
where $A$ is the fitted total MIP light yield ($S+L$) in PEU at year 0 (2010), $\tau$ is the time constant of the exponent in years, and $t$ is the year since 2010. The fit parameters are shown in table~\ref{tab:ecal_lightyield_exp_fits}.
\begin{table}[t]
\footnotesize
\centering
    \caption{Exponential fit parameters to ECal light yield at 0\,cm from the MPPC from figure~\ref{fig:ecal_distance_ly}. Reference time constants, in parentheses, from the exponential fit in  table~\ref{tab:exp_fit} are included for comparison.}
    \begin{tabular}{|c|c||c|c|c|}
    \hline
    ECal Bar Type & Readout Type & A (PEU) & $\tau$ (Ref.) (yr) & $\chi^2/\text{NDF}$ \\
    \hline
    Barrel X & Single-ended & $38.4\pm0.5$ & $45.2\pm5.1$ \quad ($47.7\pm1.1$) & $0.37/5=0.07$ \\
    Barrel Y & Single-ended & $37.2\pm0.5$ & $45.2\pm4.8$ \quad ($45.7\pm1.3$) & $1.11/5=0.22$ \\
    Barrel Z & Double-ended & $27.8\pm0.02$ & $49.5\pm3.0$ \quad ($44.1\pm0.9$) & $3.22/5=0.64$ \\
    Downstream & Double-ended & $28.0\pm0.3$ & $51.6\pm5.3$ \quad ($49.2\pm2.6$) & $2.82/5=0.56$ \\
    \hline
    \end{tabular}
    \label{tab:ecal_lightyield_exp_fits}
\end{table}

For the linear fits the annual reduction in light yield is consistent within $\sim$1$\sigma$ of the reference degradation shown in table \ref{tab:fit_param}, and similarly the time constant for the exponential fits is consistent within $\sim$1$\sigma$ of the reference values shown in table \ref{tab:exp_fit}. This suggests the ageing is dominated by the degradation of the scintillator rather than the WLS fibres.

\begin{figure}[t]
\centering
    \center{\includegraphics[width=0.49\textwidth]{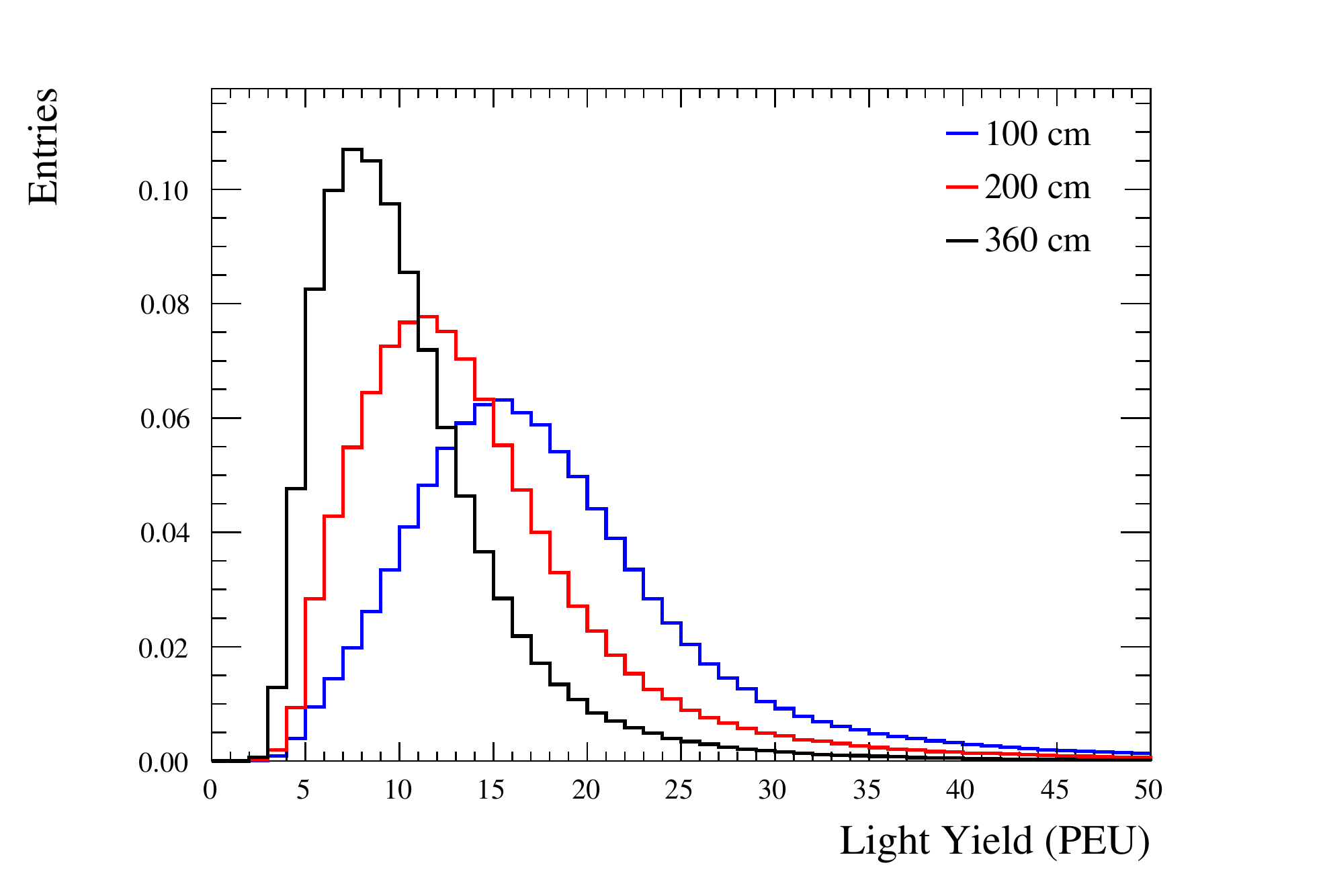}}
\caption{MIP light yield distribution in the ECal Barrel Z bars during T2K Run 9, for cosmic rays passing at distances of 100, 200 and 360~cm from the MPPCs.}
\label{fig:barrelzmipdistly}
\end{figure}

The exception to this is the Barrel Z results which lie $\sim$2$\sigma$ from the reference values and imply a slower rate of degradation than those shown in the earlier results of section \ref{sec:ageing} and \ref{sec:futureresponse}. This is likely due to some loss in MIP hit efficiency at the furthest distances from the MPPCs as the scintillator degrades. This would truncate the rising edge of the MIP light yield distribution, see for example at a distance of 360~cm in figure \ref{fig:barrelzmipdistly}, shifting the extracted MIP MPV to a slightly higher value than might be otherwise expected. The result would be an underestimate in the degradation rate extracted with this technique for the Barrel Z bars, leading to the discrepancy when making comparisons with the reference values.

Fortunately any loss in hit efficiency at the furthest distances from the MPPCs will have negligible impact on the overall hit reconstruction efficiency as the MPPC on the opposing end of the bars will continue to efficiently reconstruct these hits, as only one MPPC is required to reconstruct a hit on the double-ended readout bars.

This is confirmed by separate studies monitoring hit efficiency in the ECal modules which observed a negligible reduction ($\sim$0.1\%) in the single-hit efficiency (requiring a hit in the single-ended readout bars, or at least one hit on either end of the double-ended readout bars) across all bar types during the current lifetime of the ECal. For the double-end readout bars the double-hit efficiency (requiring a hit on both ends of a scintillator bar) has reduced by $\sim$2\% over the current lifetime.

In the future there may be some concern regarding the reconstruction of hits at the centre of the Barrel Z bars, where hits are equidistant from both sensors, and so where any impact on reconstruction efficiency will first become apparent. However, this is not a concern for the current data as shown by the MIP light yield distribution at a distance of 200~cm in figure \ref{fig:barrelzmipdistly}, but will need to be monitored.

\subsection{Fibre Degradation}

\begin{figure}[t]
\center
\begin{subfigure}{0.49\textwidth}
    \center{\includegraphics[width=\textwidth]{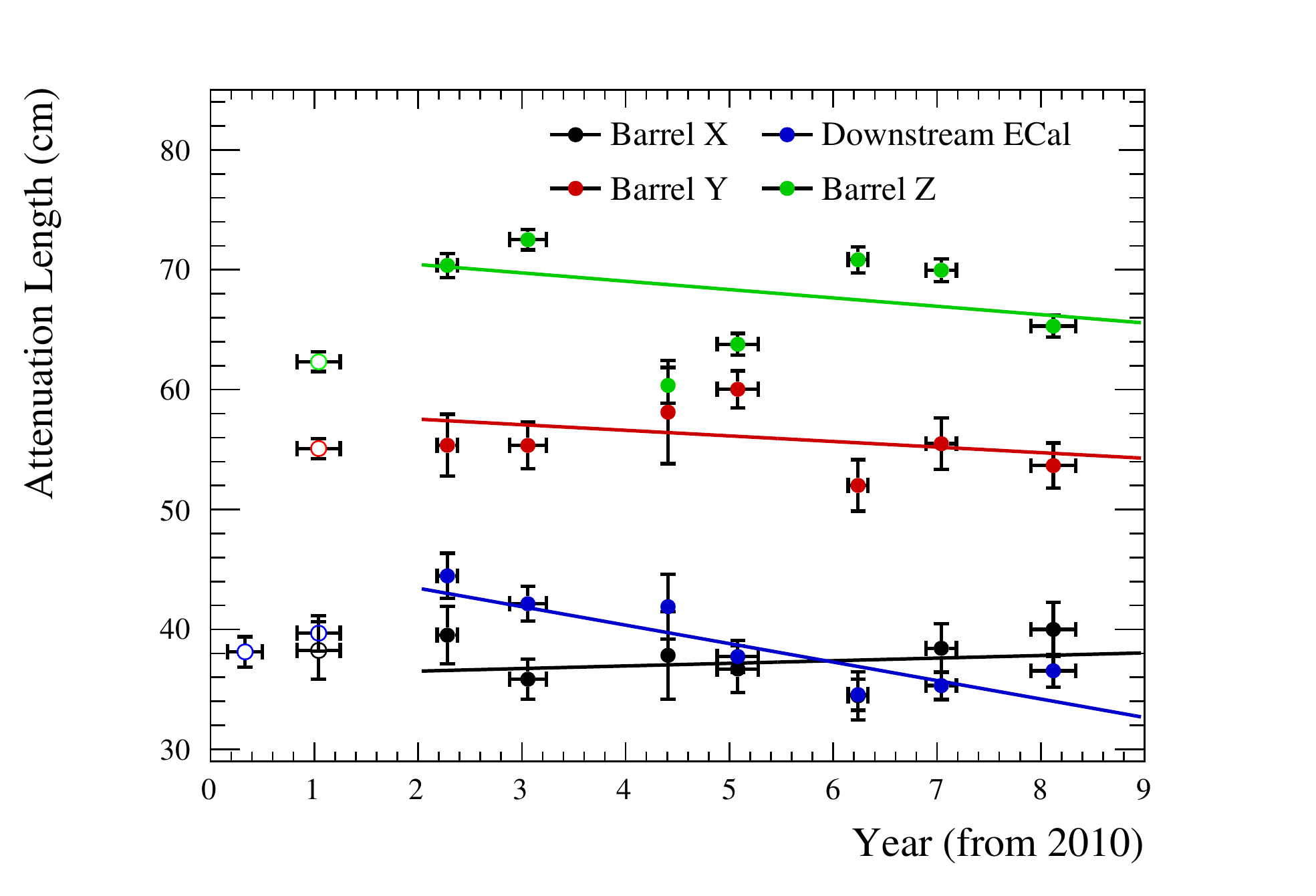}}
    \caption{Short attenuation length.}
    \label{fig:short_attenuation}
\end{subfigure}
\hfill
\begin{subfigure}{0.49\textwidth}
    \center{\includegraphics[width=\textwidth]{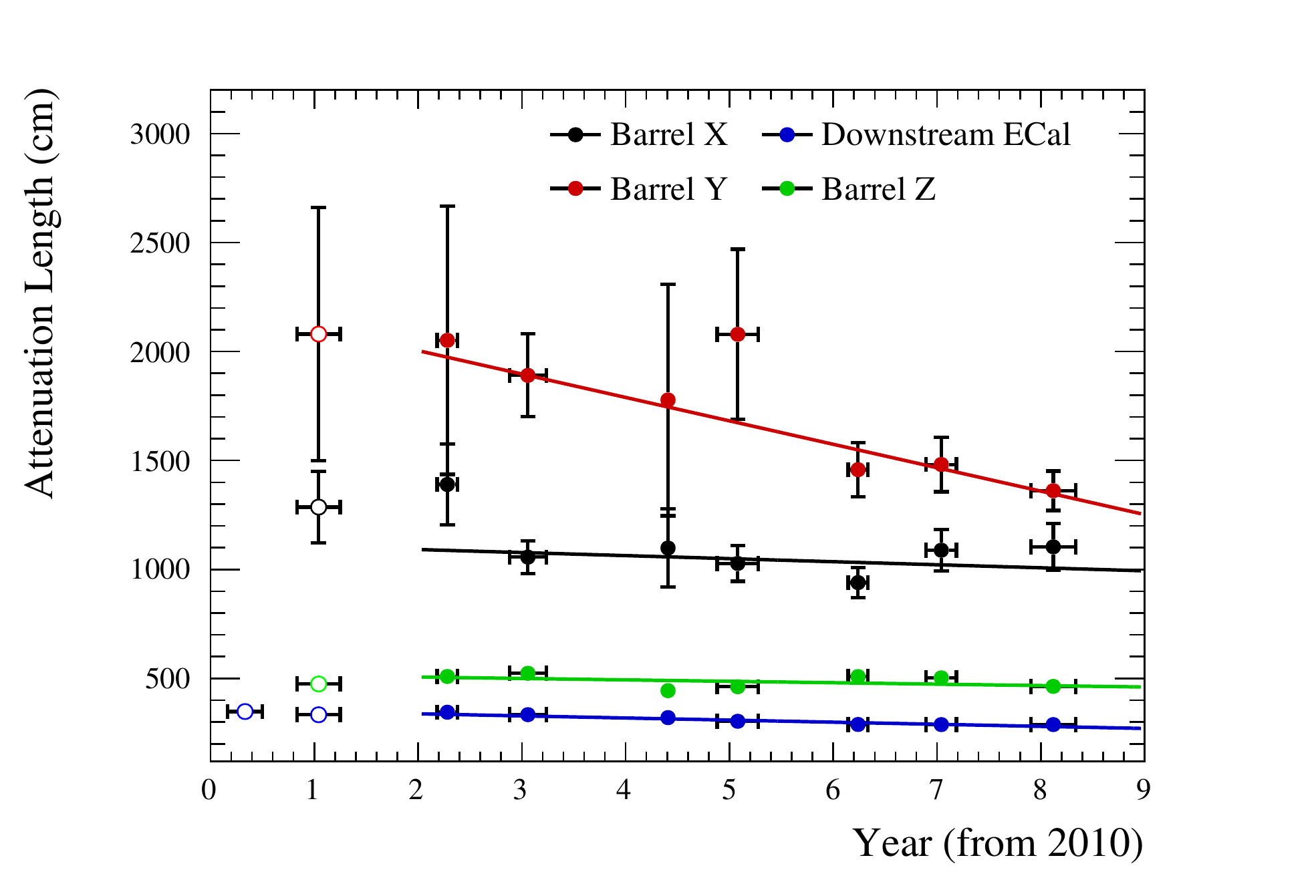}}
    \caption{Long attenuation length.}
    \label{fig:long_attenuation}
\end{subfigure}
\caption{Short and long attenuation lengths, $\lambda_S$ and $\lambda_L$ respectively, from equation \ref{eq:doubleexp}. Hollow data points are excluded from the data fits. Note the suppressed 0 for the ordinate of figure~\ref{fig:short_attenuation}.}
\label{fig:ecal_attenuation_length}
\end{figure}

Along with extracting the light yield from the fits in figure~\ref{fig:ecaldistancempv}, it is also possible to study the change in the short and long attenuation length components of the double-exponential fit, $\lambda_S$ and $\lambda_L$, respectively, for the WLS fibres. These are shown in figure~\ref{fig:ecal_attenuation_length} with linear fits applied to the data from 2012 of the form:
\begin{equation}
{\lambda_i\left(t\right)=\lambda_i\left(0\right)-k_it}\,,
\end{equation}
where $\lambda_i = \left\{\lambda_S,\lambda_L\right\}$ is the short or long attenuation length in cm at year 0 (2010), $k_i = \left\{k_S,k_L\right\}$ is the gradient of the fit in cm per year, and $t$ is the year since 2010. The fit parameters are shown in table~\ref{tab:ecal_attenuation_fits}.
\begin{table}[t]
\footnotesize
    \centering
    \caption{Linear fit parameters to ECal short and long attenuation length components of double-exponential fits from figure~\ref{fig:ecal_attenuation_length}, and the annual percentage reduction, relative to 2012 fit values.}
    \begin{tabular}{cc||c|c|c||c|}
    \cline{3-6}
     & & \multicolumn{4}{c|}{Short Attenuation Length Component} \\
    \hline
    \multicolumn{1}{|c|}{ECal Bar Type} & Readout Type & $\lambda_S\left(0\right)$ (cm) & $k_S$ (cm/yr) & $\chi^2/\text{NDF}$ & Annual Reduction (\%) \\
    \hline
    \multicolumn{1}{|c|}{Barrel X} & Single-ended & $36.1 \pm 2.3$ & $-0.22 \pm 0.41$ & $5.21/5=1.04$ & $-0.60\pm1.12$ \\
    \multicolumn{1}{|c|}{Barrel Y} & Single-ended & $58.5 \pm 2.4$ & $0.47 \pm 0.42$ & $11.10/5=2.22$ & $0.82\pm0.73$ \\
    \multicolumn{1}{|c|}{Barrel Z} & Double-ended & $71.8 \pm 1.1$ & $0.70 \pm 0.19$ & $87.16/5=17.43$ & $0.99\pm0.27$ \\
    \multicolumn{1}{|c|}{Downstream} & Double-ended & $46.4 \pm 1.8$ & $1.52 \pm 0.29$ & $8.48/5=1.70$ & $3.51\pm0.69$ \\
    \hline
    & &  \multicolumn{4}{c|}{Long Attenuation Length Component} \\
    \hline
    \multicolumn{1}{|c|}{ECal Bar Type} & Readout Type & $\lambda_L\left(0\right)$ (cm) & $k_L$ (cm/yr) & $\chi^2/\text{NDF}$ & Annual Reduction (\%) \\
    \hline
    \multicolumn{1}{|c|}{Barrel X} & Single-ended & $1119 \pm 117$ & $13.9 \pm 19.8$ & $5.98/5 = 1.20$ & $1.27\pm1.82$ \\
    \multicolumn{1}{|c|}{Barrel Y} & Single-ended & $2218 \pm 262$ & $107.2 \pm 35.6$ & $1.68/5 = 0.34$ & $5.35\pm1.92$ \\
    \multicolumn{1}{|c|}{Barrel Z} & Double-ended & $520 \pm 6$ & $6.6 \pm 1.1$ & $146.47/5 = 29.29$ & $1.30\pm0.22$ \\
    \multicolumn{1}{|c|}{Downstream} & Double-ended & $354 \pm 7$ & $9.1 \pm 1.1$ & $13.29/5 = 2.67$ & $2.71\pm0.33$ \\
    \hline
    \end{tabular}
    \label{tab:ecal_attenuation_fits}
\end{table}

The short attenuation length varies between 36 and 72\,cm, increasing as the bar length increases, and it appears to be consistent with minimally $\left(<1\%\right)$ or not degrading with time. The exception to this is the Downstream ECal which shows a higher degradation rate of $3.51\pm0.69\%$, although if the earlier Run 1 and 2 data were to be included this would substantially reduce.

For the long attenuation length, the single-ended (mirrored) bars have much longer attenuation lengths compared to the double-ended readout bars, $\sim$1120 and $\sim$2220~cm for the Barrel X and Y bars, respectively, compared to $\sim$520 and $\sim$355~cm for the Barrel Z and Downstream bars, respectively. This discrepancy is due to the mirrored bars having two signals, direct transmission down the WLS fibres to the MPPCs, and reflected transmission, the combination of which is not accounted for in the fits, and so these are not true measurements of the long attenuation length.

For comparison, early fibre scanning work during construction on the ND280 ECals measured short and long attenuation lengths for the WLS fibres in the range 21--31~cm and 390--410~cm, respectively \cite{Allan:2013ofa}.

Kuraray have also measured the attenuation length of their fibres from light yield measurements over a distance range of 100--300~cm, fitting the distribution with a single exponential function and extracting an attenuation length of $>$350~cm \cite{WLS}, in agreement with our long attenuation length results.

The Mu2e collaboration which also uses Kuraray Y\nobreakdash-11 WLS fibres has measured the attenuation length of the fibres, but over substantially longer fibre lengths of 25~m. In a 2015 study they applied a double-exponential fit to their data of the same form shown in equation \ref{eq:doubleexp} and extracted short and long attenuation lengths of 4.76 and 9.02~m, respectively \cite{DeZoort:2015pfn}. A later study in 2018 separated the data into two independent exponential fits over the ranges 0.5--3.0~m and 3.0--25.0~m, and again extracting short and long attenuation lengths, this time of $5.1\pm0.2$ and $8.2\pm0.1$~m, respectively \cite{osti_1580729}. In both cases their short attenuation length measurement is consistent with the (double-ended readout bars) long attenuation lengths we have obtained. Perhaps of greater interest though are their measurements of attenuation length as a function of wavelength which show very short attenuation lengths of less than 50~cm at 490~nm, approaching the peak quantum efficiency for our MPPCs which occurs at 440~nm \cite{MPPCTN} (unfortunately the Mu2e measurements do not extend to wavelengths below 490~nm) and longer attenuation lengths of $\sim$400~cm at 510~nm.

Kuraray Y\nobreakdash-11 WLS fibres absorb light over wavelengths of $\sim$360--490~nm, and emit between $\sim$460--570~nm \cite{WLS}. Our two attenuation length measurements can then be readily explained. A short attenuation length attributed to the overlapping absorption and emission regions of the Y\nobreakdash-11 WLS fibres around $\sim$475~nm (near the maximum quantum efficiency of the MPPCs), and a longer attenuation length coinciding with the emission only region of the Y\nobreakdash-11 WLS fibres at $>$490~nm (mean emission value of $\sim$510~nm \cite{Allan:2013ofa}), as corroborated by the single wavelength Mu2e attenuation length measurements.

Irrespective of the mirroring or not, the long attenuation lengths do appear to be degrading by between $1.27\%$ and $5.35\%$ per year, although the single-ended (mirrored) bars have significant uncertainties on those rates.

As to why the long attenuation length would show degradation whilst the short attenuation length does not is unknown. Potentially a wavelength dependent change in the opacity of the fibres has occurred, allowing shorter wavelengths\textbf{} ($<$490~nm) to propagate in a consistent manner over the current lifetime of the WLS fibres, whilst increasing the opacity to longer wavelengths ($>$490~nm). However this is purely conjecture and we cannot ascribe a mechanism for such behaviour.

\section{Conclusions}
The rate of ageing for the different scintillator subsystems of ND280 and INGRID has been studied. The materially identical ECal, P\O D, FGD and INGRID observe an annual deterioration in the light yield of 1.2--2.2\%, whilst the SMRD shows a somewhat lower rate of degradation at $0.9\pm0.4$\%. These results are comparable to similar studies by the MINOS experiment ($\sim$2\%)~\cite{Michael:2008bc}, but inconsistent with a shorter duration study undertaken by the MINER$\nu$A experiment ($\sim$7.5\%)~\cite{Aliaga:2013uqz}, both of which use scintillator bars which are materially identical to the majority of T2K subsystems.

Modelling the decrease in light yield of the ECal as an exponential shows that the response is expected to halve by 2040, at which time the reduced response may become challenging for some physics analyses. This may be beyond the lifetime of the ND280 detector, but if its use continues into the Hyper-Kamiokande era then it motivates the development of improved detector simulation and reconstruction algorithms to mitigate the impact. The higher initial response and lower degradation rates of the other ND280 subsystems and INGRID implies their physics capabilities are less likely to be negatively impacted over the same timescales.

The additional study to disentangle the degradation of the scintillator and WLS fibres within the ECal shows that the majority of the ageing can be attributed to the degradation of the scintillator rather than the WLS fibres. The short component of the WLS fibre attenuation length appears consistent with not degrading, although the long attenuation component does appear to degrade by between 1--5\% per year, and the cause of this apparent discrepancy is unknown.

A summary of the annual light yield reduction for each ND280 scintillator detector can be found in table~\ref{summary:table}.
\begin{table}[t]
    \footnotesize
    \centering
    \caption{Summary of the annual light yield reduction for ND280 and INGRID subsystems, relative to their 2012 light yields.}
    \begin{tabular}{|c|c|}
    \hline
    Subsystem & Annual Light Yield Reduction (\%) \\
    \hline
    P\O D & $1.8\pm0.2$ \\
    FGD & $1.2\pm0.2$ \\
    ECal & $\left(1.9-2.2\right)\pm0.1$ \\
    SMRD & $0.9\pm0.4$ \\
    INGRID & $1.8\pm0.1$ \\
    \hline
    \end{tabular}
    \label{summary:table}
\end{table}

\vspace{\baselineskip}
{\setlength{\parindent}{0cm}Data supporting the results reported in this paper are openly available from the Zendo data repository and can be found here \cite{zenodo}.}

\acknowledgments
We thank the J-PARC staff for superb accelerator performance. We thank the CERN NA61/SHINE Collaboration for providing valuable particle production data. We acknowledge the support of MEXT, JSPS KAKENHI (JP16H06288, JP18K03682, JP18H03701, JP18H05537, JP19J01119, JP19J22440, JP19J22258, JP20H00162, JP20H00149, JP20J20304) and bilateral programs\linebreak (JPJSBP120204806, JPJSBP120209601), Japan; NSERC, the NRC, and CFI, Canada; the CEA and CNRS/IN2P3, France; the DFG (RO 3625/2), Germany; the INFN, Italy; the Ministry of Education and Science(DIR/WK/2017/05) and the National Science Centre (UMO-2018/30/E/ST2/00441), Poland; the RSF19-12-00325, RSF22-12-00358 and the Ministry of Science and Higher Education (075-15-2020-778), Russia; MICINN (SEV-2016-0588, PID2019-107564GB-I00, PGC2018-099388-BI00) and ERDF funds and CERCA program, Spain; the SNSF and SERI (200021\_185012, 200020\_188533, 20FL21\_186178I), Switzerland; the STFC, UK; and the DOE, USA. We also thank CERN for the UA1/NOMAD magnet, DESY for the HERA-B magnet mover system, NII for SINET5, the WestGrid and SciNet consortia in Compute Canada, and GridPP in the United Kingdom. In addition, the participation of individual researchers and institutions has been further supported by funds from the ERC (FP7), “la Caixa” Foundation (ID 100010434, fellowship code LCF/BQ/IN17/11620050), the European Union’s Horizon 2020 Research and Innovation Programme under the Marie Sklodowska-Curie grant agreement numbers 713673 and 754496, and H2020 grant numbers RISE-GA822070-JENNIFER2 2020 and RISE-GA872549-SK2HK; the JSPS, Japan; the Royal Society, UK; French ANR grant number ANR-19-CE31-0001; and the DOE Early Career programme, USA.

\bibliographystyle{ieeetr}
\bibliography{bibliography}
\end{document}